\documentclass[aps,prd,twocolumn,a4paper,superscriptaddress]{revtex4-1}
\usepackage{graphicx}
\usepackage{multirow}
\usepackage{latexsym}
\usepackage{hyperref}
\usepackage[british]{babel}
\usepackage{amsmath}

\begin{document}

\newcommand{\be}{\begin{equation}}
\newcommand{\ee}{\end{equation}}
\newcommand{\bq}{\begin{eqnarray}}
\newcommand{\eq}{\end{eqnarray}}
\newcommand{\bsq}{\begin{subequations}}
\newcommand{\esq}{\end{subequations}}
\newcommand{\bc}{\begin{center}}
\newcommand{\ec}{\end{center}}

\title{Dynamics of junctions and the multi-tension velocity-dependent one-scale model}

\author{I. Yu. Rybak}
\email[]{Ivan.Rybak@astro.up.pt}
\affiliation{Centro de Astrof\'{\i}sica da Universidade do Porto, Rua das Estrelas, 4150-762 Porto, Portugal}
\affiliation{Instituto de Astrof\'{\i}sica e Ci\^encias do Espa\c co, CAUP, Rua das Estrelas, 4150-762 Porto, Portugal}
\affiliation{Faculdade de Ci\^encias, Universidade do Porto, Rua do Campo Alegre 687, 4169-007 Porto, Portugal}
\author{A. Avgoustidis}
\email[]{Anastasios.Avgoustidis@nottingham.ac.uk}
\affiliation{School of Physics and Astronomy, University of Nottingham, University Park, Nottingham NG7 2RD, United Kingdom}
\author{C. J. A. P. Martins}
\email{Carlos.Martins@astro.up.pt}
\affiliation{Centro de Astrof\'{\i}sica da Universidade do Porto, Rua das Estrelas, 4150-762 Porto, Portugal}
\affiliation{Instituto de Astrof\'{\i}sica e Ci\^encias do Espa\c co, CAUP, Rua das Estrelas, 4150-762 Porto, Portugal}

\date{11 December 2018}

\begin{abstract}
The dynamics of string junctions and their influence on the evolution of cosmic superstring networks are studied in full detail. We review kinematic constraints for colliding strings in a Friedmann-Lema\^itre-Robertson-Walker background and obtain the average distribution of possible string configurations after string collisions. The study of small-scale structure enables us to investigate the average growth/reduction rate of string junctions for a given cosmic string network. Incorporating the averaged junction dynamics into the velocity-dependent one-scale model for multi-tension string networks, we improve the semi-analytic description and quantitative understanding of cosmic superstring network evolution.
\end{abstract}
\maketitle

\section{Introduction} \label{Introduction}

A possible outcome of brane inflation is the production of a network of cosmic superstrings \cite{BurgessMajumdarNolteQuevedoRajeshZhang,  SarangiTye,  JonesStoicaTye, DvaliKalloshProeyen, DvaliVilenkin, PolchinskiCopelandMyers, FirouzjahiTye}. This particle physics relic offers an exciting observational opportunity to shed light on high energy physics processes that may have taken place in the earliest phases of the evolution of our universe. In order to have an accurate prediction for the observational signals from cosmic superstrings, the cosmological evolution of their networks must be studied rigorously. Cosmic superstrings can in general have rather complicated properties compared to conventional cosmic strings (for reviews of ordinary cosmic strings see \cite{HindmarshKibble, ShellardVilenkin}). It is then anticipated that these additional features bring about important changes to the network evolution and should be carefully taken into account. 

Cosmic superstring collisions give rise to string junctions, and the corresponding multi-tension string networks are very challenging to model numerically. Generically, the networks will not be Brownian, but will be made up of a series of segments with different tensions and lengths, connected at junctions. Tensions of cosmic superstrings are related by the BPS conditions \cite{WITTEN1996, Schwarz, PolchinskiCopelandMyers}
\begin{equation}
\label{PQtensions}
\mu_{pq} = \mu_F \sqrt{p^2 + q^2/g_s^2},
\end{equation}
where $p$ and $q$ are integers that represent the bound state of $p$ $F$-strings and $q$ $D$-strings (the Ramond-Ramond scalar is set to zero $C_0=0$). Thus the network will contain a hierarchy of string tensions, and the dynamical equations describing the density and characteristic length and velocity of each type will all be coupled to each other.

Lattice simulations for such networks are prohibitively time-consuming, so that network evolution cannot be reproduced with both high-resolution and sufficiently large dynamical ranges. Furthermore, superstring network evolution can only be modelled effectively from numerical simulations: not all properties of cosmic superstring networks can be captured by a field theory  \cite{HindmarshSaffin, Saffin, RajantieSakellariadouStoica, UrrestillaVilenkin, LizarragaUrrestilla}. Given these computational challenges, a more fruitful approach to this problem is through semi-analytic methods---a generalized Velocity-dependent One Scale (VOS) model \cite{MartinsShellard,MartinsShellard2,Book}. 

To model the cosmological evolution of cosmic superstring networks one needs to include the important features characterising their complex interactions. In particular, the quantum description of string interactions \cite{POLCHINSKI, DaiPolchinski, JacksonJonesPolchinski, Jackson, HananyHashimoto}, and the kinematic constraints on junction production \cite{CopelandKibbleSteer,CopelandKibbleSteer2,CopelandFirouzjahiKibbleSteer, SalmiAchucarroCopelandKibblePutterSteer, BevisSaffin} have been included in a VOS description. Exploration of these analytic models indicates that these complex networks can still achieve scaling behaviour \cite{TyeWassermanWyman, AvgoustidisShellard, AvgoustidisCopeland,  PourtsidouAvgoustidisCopelandPogosianSteer}, although the corresponding scaling densities and velocities may be different from those of ordinary strings. Here we obtain the generalisation of the relevant kinematic conditions on junction production for the case of a Friedmann-Lema\^itre-Robertson-Walker (FLRW) metric \cite{FirouzjahiKhoeini-MoghaddamKhosravi}. This generalisation is probed through an example of collisions of straight strings in an FLRW background, which provides a time-dependent ``angle-velocity" diagram for junction production processes (similar to the one obtained for Minkowski space in \cite{CopelandKibbleSteer2}). We further model, for the first time, the average evolution of string junctions within the network.

In order to model the evolution of junctions in the string network we first review the description of small-scale structure found in  \cite{PolchinskiRocha,KibbleCopeland2}. This model---which we extend in the present work---provides a useful tool for describing the average growth/reduction of junctions in a cosmic superstring network. The derived average change of string lengths due to junction evolution leads to a generalization of the VOS model describing superstring network evolution with dynamical junctions (Section \ref{VOS model evolution with dynamcial junctions}). This work directly builds on previous extensions of the VOS model \cite{MartinsShellard,MartinsShellard2} for describing superstring networks, which can be found in \cite{TyeWassermanWyman, AvgoustidisShellard, PourtsidouAvgoustidisCopelandPogosianSteer}. We briefly discuss some of the solutions of our extended model, and compare them with earlier results.

\section{Junction dynamics} \label{Junction dynamics}

Let us start our consideration of junction dynamics from the action for three connected strings, which can be written as \cite{Sharov1997, Hooft}
\begin{equation}
\begin{gathered}
   \label{ActionJunction}
   S = -\sum_i \mu_i \int \Theta(s_i(\tau)-\sigma_i) \sqrt{-\gamma_i} d \sigma_i d \tau + \\
   +\sum_i \int \mathrm{f}_{i \mu} \left( x_i^{\mu}(s_i(\tau),\tau) - \mathcal{X}^{\mu} \right) d \tau,
\end{gathered}  
\end{equation}
where the index $i=1,2,3$ specifies one of the three string segments (there is no summation over these repeated indices -- summation happens only if the symbol $\sum$ is written explicitly), $\mu_i$ are the string tensions, $\sigma_i$ and $\tau$ are the spacelike and timelike coordinates on the string worldsheet(s), $\gamma_{iab}$ are the worldsheet metrics with determinants $\gamma_i$, $\mathrm{f}_{i \mu}$ are Lagrange multipliers, $\Theta$ is the Heaviside step function, $\mathcal{X}^{\mu}$ is the space-time position of the vertex (where all three strings are connected), and $s_i$ is the parametrisation value of $\sigma_i$ at the vertex $\mathcal{X}^{\mu}$.

For each of the strings, we use the following worldsheet parametrization (dropping the index $i$ for simplicity):
\begin{equation}
\begin{split}
   \label{Parametrization}
& \sigma^0 = \tau , \quad \sigma^1 = \sigma, \\
& \; \; g_{\mu \nu} \, \dot{x}^{\nu} x^{\prime \, \mu} = 0 \,, 
\end{split}
\end{equation}
where $g_{\mu \nu}$ is the background spacetime metric, and $\tau$ is the (conformal) background time coordinate. The dot and prime denote differentiation with respect to $\tau$ and $\sigma$ respectively. We are working in a flat FLRW metric
\begin{equation}
\label{FLRW}
ds^2 = a(\tau)^2 \left( \tau^2 - dl^2 \right),
\end{equation}
where $a$ is a scale factor and $dl^2$ is the line element of flat spatial sections.  The conformal time $\tau$ is related to physical time $t$ by $a d \tau = d t$. We shall now study the equations of motion describing the dynamics of this 3-string system in FLRW geometry.  

By variation of the action (\ref{ActionJunction}) with respect to $\textbf{x}_{i}$, one can obtain the standard string equations of motion 
\begin{equation}
\label{MicroEqOfMotString}
\begin{split}
& \qquad \dot{\varepsilon} + 2 \varepsilon \frac{\dot{a}}{a} \dot{\textbf{x}}^2 = 0 ,\\
& \ddot{\textbf{x}} + 2 \frac{\dot{a}}{a} \dot{\textbf{x}} (1-\dot{\textbf{x}}^2) = \frac{1}{\varepsilon} \left( \frac{\textbf{x}^{\prime}}{\varepsilon} \right)^{\prime},
\end{split}
\end{equation} 
where $\textbf{x}$ is the three-dimensional spatial vector, $\dot{\textbf{x}} \equiv \frac{\partial \textbf{x}}{\partial \tau} $, $\textbf{x}^{\prime} \equiv \frac{\partial \textbf{x}}{\partial \sigma} $ and $\varepsilon^2 = \frac{\textbf{x}^{\prime \, 2}}{1-\dot{\textbf{x}}^2} $.

An additional equation from the $\textbf{x}_{i}$ variation appears due to the boundary term in the action (\ref{ActionJunction}) and has the form \cite{Pourtsidouphdthesis}
\begin{equation}
\label{Boundary}
\mu_i \left( \frac{\textbf{x}_i^{\prime}}{\varepsilon_i} + \varepsilon_i \dot{s}_i \dot{\textbf{x}}_i \right) = \textbf{f}_i,
\end{equation}
on the string junction, i.e. all functions in (\ref{Boundary}) are evaluated at $(s(\tau),\tau)$. Further, from the variation with respect to $\mathcal{X}^{\mu}$ it follows that the Lagrange multipliers satisfy the condition 
\begin{equation}
\label{ForceBalanceEq}
\sum_i \textbf{f}_i = 0 \, ,
\end{equation}
again evaluated on the junction. 

Finally, from the variation of the Lagrange multipliers $\mathrm{f}_i$ we obtain the previously mentioned conditions that $\sigma=s$ at the vertex,
\begin{equation}
\label{VertexEq}
\textbf{x}_i(s_i(\tau),\tau)=\textbf{X}(\tau),
\end{equation}
where $\textbf{X}$ is the spatial part of $\mathcal{X}^{\mu}$.

In the case of Minkowski space, which corresponds to $\varepsilon=1$, there is an exact solution of the equations of motion (\ref{MicroEqOfMotString}). This solution allows one to pick out right/left moving modes on each string and, from those, obtain the dynamics of the junction  \cite{CopelandKibbleSteer}. For the expanding FLRW metric there is no general solution of the string equations (\ref{MicroEqOfMotString}). However, it is possible to build a convenient analogue of right/left moving modes (this approach for the FLRW metric was used in \cite{FirouzjahiKhoeini-MoghaddamKhosravi} for the analysis of loops with junctions):

\begin{equation}
 \begin{split}
\label{UnitVectorsExpandUni}
 & \textbf{q} = \textbf{x}^{\prime}/\varepsilon+\dot{\textbf{x}}, \\
 & \textbf{p} = \textbf{x}^{\prime}/\varepsilon-\dot{\textbf{x}}.
\end{split}
\end{equation}

In the case of monotonic expansion of the scale factor in the FLRW metric, it is reasonable to assume that ingoing and outgoing waves are distinguishable and won't be mixed during the evolution. That is why we can assume that vectors $\textbf{q}$ and $\textbf{p}$ are outgoing and ingoing waves, respectively, for the vertex $\textbf{X}$ in an expanding universe.

Taking into account the fact that $|\textbf{q}|^2=1$, $|\textbf{p}|^2=1$ and using equations (\ref{MicroEqOfMotString})-(\ref{VertexEq}) one can carry out an analysis which generalizes the one done for Minkowski space in~\cite{CopelandKibbleSteer}. This leads to the following equations for the 3-string junction in an expanding universe
\begin{equation}
 \begin{split}
\label{JunctionsUnit1}
  & \sum_i \mu_i \left( (\textbf{q}_i+\textbf{p}_i) + \varepsilon_i \dot{s}_i (\textbf{q}_i-\textbf{p}_i) \right) = 0.
\end{split}
\end{equation}
\begin{equation}
 \begin{split}
\label{JunctionsUnit2}
  & \dot{\textbf{X}}=-\frac{1}{\mu} \sum_i \mu_i (1-\varepsilon_i \dot{s}_i)\textbf{p}_i,
\end{split}
\end{equation}
\begin{equation}
 \begin{split}
\label{JunctionsUnit3}
  & \frac{(1-\varepsilon_i \dot{s}_i) \mu_i}{\mu}=\frac{M_i (1-c_i)}{ \mathcal{M}},
\end{split}
\end{equation}
where we have defined
\begin{equation}
\begin{gathered}
\label{c_i}
c_i = \frac{1}{2} \sum_{jk} | \epsilon_{ijk} |\, \textbf{p}_j \! \cdot \! \textbf{p}_k \, , \\ 
M_i = \mu_i^2 - \frac{1}{2} \sum_{jk} (\epsilon_{ijk} (\mu_j-\mu_k))^2, \\
\mathcal{M}=\sum_i M_i (1-c_i),\\
\mu = \sum_i \mu_i,
\end{gathered}
\end{equation}
and $\epsilon_{ijk}$ is the Levi-Civita symbol.

One then observes that the equations describing string junctions in an expanding universe can be obtained from those in Minkowski space by just changing $\dot{s}_i$ to $\varepsilon_i \dot{s}_i$ and the definitions of outgoing and ingoing modes to the generalised quantities (\ref{UnitVectorsExpandUni}). The new set of equations (\ref{JunctionsUnit3}) tell us how growth/reduction of string length proceeds in  an expanding FLRW universe. An important issue that stems from the above treatment is to understand under which conditions junction production can successfully take place. The condition for junction formation just requires $\dot{s}_3>0$ and will be considered for straight strings below.

\subsection{Solution for a straight string in a FLRW background} \label{Solution for a straight string in FLRW background}

It is always possible to choose a sufficiently small region, near the collision point, where strings can be considered straight. Hence, to understand the conditions under which the collision of strings leads to junction production, it suffices to consider straight string collisions. For this purpose we first need to construct a proper straight string solution in a FLRW metric. Let's consider the following straight string ansatz, similar to the case of Minkowski space,
\begin{equation}
\begin{split}
\label{ansatzFLRW}
 \textbf{x} = \left\lbrace C_1 \sigma \cos \alpha; \, C_1\sigma \sin \alpha; \, F(\tau) \right\rbrace,
 \end{split}
\end{equation}
where $C_1$ is a constant that will be defined later and the function $F(\tau)$ is to be determined by the dynamics. 

The ansatz (\ref{ansatzFLRW}) describes a straight string located on the $XY$-plane and moving along the $Z$-axis. The angle $\alpha$ defines the orientation of the string in the $XY$-plane with respect to the $X$-axis.

It should be noted that while the dependence of this FLRW string solution on the spacelike worldsheet coordinate $\sigma$ is the same as in Minkowski space, the physical meaning of $\sigma$ in these two cases is not identical. Any interval of string in Minkowski space has a fixed length, which does not depend on time, while in a FLRW metric the length of a chosen string interval is stretching as time evolves, which implies a continuous ``effective" reparametrization of the string ansatz (\ref{ansatzFLRW}). Meanwhile, the time component $\tau$ is substituted by the function $F(\tau)$, due to the absence of time translational invariance.

Using the form (\ref{ansatzFLRW}) we can obtain the following useful quantities
\begin{subequations}
\begin{align}
\label{prime}
 &  \textbf{x}^{\prime} = \left\lbrace C_1 \cos \alpha; \, C_1 \sin \alpha; \, 0 \right\rbrace, \\
\label{dot}
 &  \qquad \dot{\textbf{x}} = \left\lbrace 0; \, 0; \, \dot{F}(\tau) \right\rbrace, \\
 \label{eps}
 & \,  \varepsilon = \sqrt{\frac{\textbf{x}^{\prime \, 2}}{1-\dot{\textbf{x}}^2}} = \sqrt{\frac{C_1^2}{1-\dot{F}^2}},\\
  \label{epsdot}
 &  \quad \; \; \dot{\varepsilon} = |C_1| \frac{\dot{F}\ddot{F}}{\left( 1-\dot{F}^2 \right)^{3/2}}.
\end{align}
\end{subequations}

Using the expressions~(\ref{prime})-(\ref{epsdot}), one can show that the equations of motion (\ref{MicroEqOfMotString}) are reduced to a single equation for the function $F(\tau)$
\begin{equation}
\begin{split}
\label{Ffunc}
 \ddot{F}+2\frac{\dot{a}}{a} \dot{F}(1-\dot{F}^2)=0.
 \end{split}
\end{equation}

When the scale factor is $a \propto \tau^{n}$, equation~(\ref{Ffunc}) has an exact solution 
\begin{equation}
\begin{split}
\label{Solution}
 F(\tau,v)= \pm \tau \, \; _{2}F_{1}\left( \frac{1}{2}; \, \frac{1}{4 n}; \, \frac{1+4 n}{4 n}; \, -\left(\frac{\tau^{2 n}}{\gamma_v v} \right)^2  \right),
 \end{split}
\end{equation}
where $_{2}F_{1}$ is the Gauss hypergeometric function, $\gamma_{v} = 1/ \sqrt{1-v^2}$ is the Lorentz factor, and we introduced the factor $\left(\gamma_v v\right)^{-2}$ to recover the Minkowski solution for $n \rightarrow 0$.

The solution (\ref{Solution}) is monotonic in the argument $\tau$ for all positive $n$, which is anticipated from the physical interpretation of (\ref{Solution}). The string has initial velocity $(0,0,v)$ and initial position $Z=0$ at $t=0$, cf. (\ref{ansatzFLRW}). The final vector solution of a straight string in FLRW metric can be written as
\begin{equation}
\begin{split}
\label{SolutionFinal}
 &\textbf{x} =\left\lbrace  \sigma \cos \alpha; \, \sigma \sin \alpha; \, F(\tau, v) \right\rbrace,
 \end{split}
\end{equation}
where $C_1$ was chosen to be $1$.

It is seen that the limit $n \rightarrow 0$ does not reproduce $\varepsilon=1$, which was chosen by parametrisation in \cite{CopelandKibbleSteer}. The use of the vectors (\ref{UnitVectorsExpandUni}) allows us to employ any other worldsheet parametrization for $\sigma$ (with $\varepsilon \neq 1$) without affecting the final result. We have thus chosen the parametrization (\ref{SolutionFinal}) for convenience.

\subsection{Collision of straight strings in expanding universe} \label{Collision of straight strings in expanding universe}

In the above we outlined our approach for the description of connected strings in the FLRW metric and constructed one simple straight string solution. Consequently, we can consider the collision of straight strings in a FLRW background and investigate under which conditions a junction can be produced. We are going to consider only the situation when the scale factor behaves as $a \propto \tau^{n}$. 

Let us model the situation when two strings $\textbf{x}_{1,2}$ are moving towards each other along the $Z$ axis, oriented at angles $(\pm \alpha)$ to the $X$-axis on the $XY$-plane, and having equal initial speeds $\upsilon$.  The third string $\textbf{x}_3$ is created in the $XY$-plane, with initial velocity $u$ and orientation defined by an angle $\theta$ to the $X$-axis
\begin{equation}
\begin{split}
\label{ThreeStrings}
 & \textbf{x}_{1,2} = \left\lbrace -\sigma \cos \alpha; \,\mp \sigma \sin \alpha; \, \pm F(\tau, v) \right\rbrace, \\
  & \quad \; \textbf{x}_{3} = \left\lbrace \sigma \cos \theta; \, \sigma \sin \theta; \,  F(\tau,u) \right\rbrace.
 \end{split}
\end{equation}

In order to have a simple analytic comparison for the expanding FLRW metric and Minkowski space, let us consider the case when the tension of the first string is the same as the tension of the second, $\mu_1=\mu_2$. This implies $u=0$ and $\theta=0$. Using the equality (\ref{JunctionsUnit3}), one can show that the solution for $\dot{s}_3$ is

\begin{equation}
\begin{split}
\label{s3}
 &  \dot{s}_3=\frac{2 \mu_1 \tilde{\gamma}(\upsilon, \tau) \cos \alpha -\mu_3}{2 \mu_1 - \mu_3 \tilde{\gamma}(\upsilon, \tau) \cos \alpha}, 
 \end{split}
\end{equation}
where $\tilde{\gamma}(\upsilon, \tau) = \frac{\gamma_{\upsilon}^{-1}}{\sqrt{1+\upsilon^2 \left( \tau^{-4 n} - 1 \right)}}$.

\begin{figure}
\centering
\includegraphics[width=0.47\textwidth]{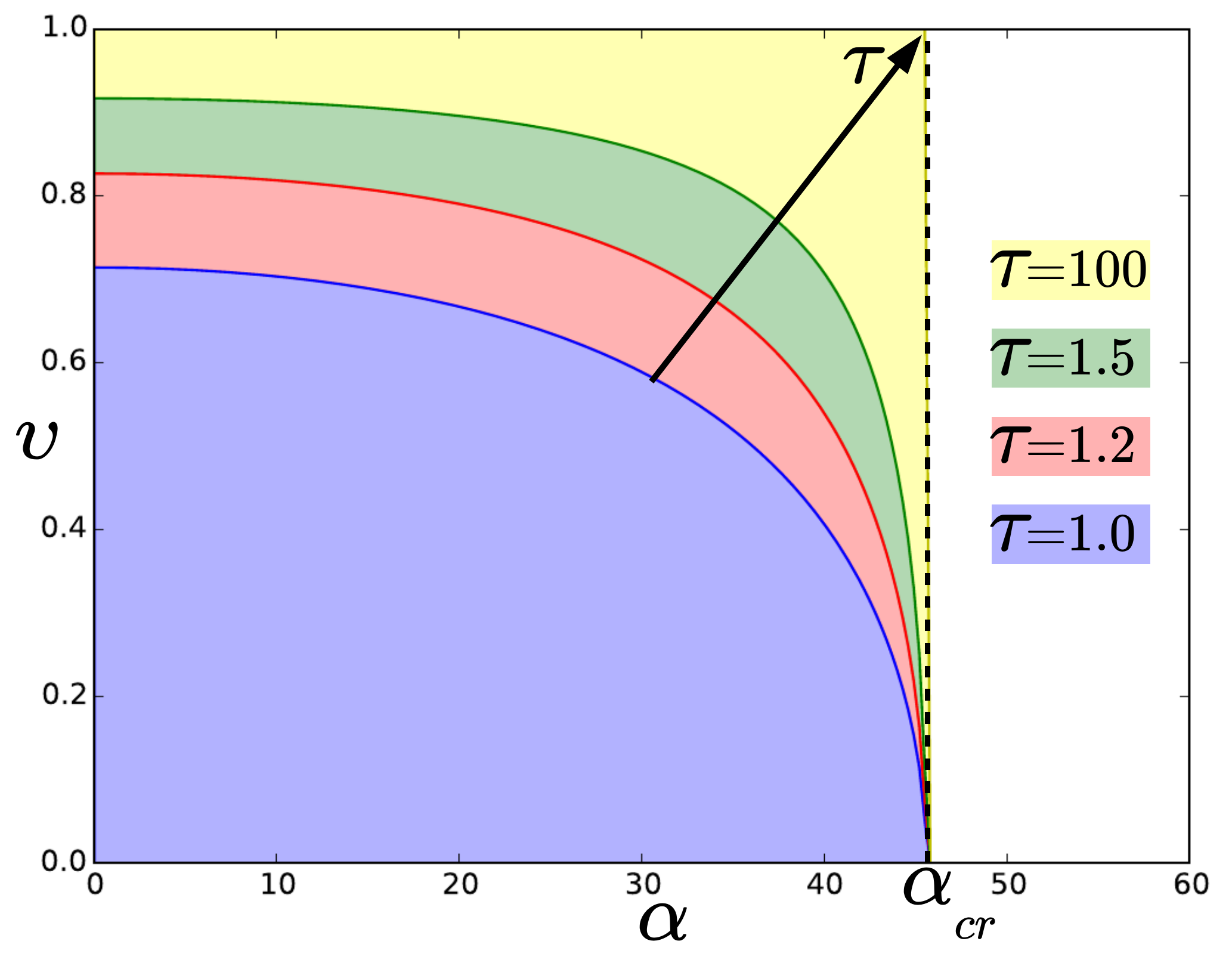}
\caption{\label{fig:areas} Range of parameters: ``initial velocity" $\upsilon$ and angle $\alpha$, which allow junction production ($\dot{s}_3>0$) for the case when the heaviest string has the tension $\mu_3=1.4 \, \mu_1=1.4 \, \mu_2$. The strings are evolving in an expanding FLRW metric with $a \propto \tau^{n}$, and the plot corresponds to the specific case $n=1.0$ (radiation era). The first (blue) area corresponds to the moment $\tau=1.0$; later evolution is represented by the other colours, and the full area (yellow), corresponding to $\alpha_{\text{cr}}$, is reached in the limit $\tau \rightarrow \infty$.}
\end{figure}

Comparing equation (\ref{s3}) with the corresponding result for Minkowski space in \cite{CopelandKibbleSteer}, it is seen that the Lorentz factor $\gamma_{\upsilon}^{-1}$ is substituted by the function $\tilde{\gamma}$, which approaches $\gamma_{\upsilon}^{-1}$ when $n\rightarrow0$, or when $\tau=1$. Let us build the region in velocity-angle space, when the junction can be produced ($\dot{s}_3>0$) - see figure \ref{fig:areas}. The value $\tau=1$ corresponds to the initial conditions of the colliding strings, which also coincides with the solution for Minkowski space. It is seen that as $\tau$ grows the region where $\dot{s}_3>0$ is enlarged. This effect appears due to the string velocity decreasing in the expanding universe. Eventually the string speed approaches zero and a junction can be formed for all possible collisions at angles smaller than critical angle $\alpha_{cr} = \arccos \left( \mu_3/(2 \mu_1) \right)$.

As a result, junction production for straight strings in an FLRW universe is different from the Minkowski case only by a change of the relative string velocity. This result suggests that we can apply the approach developed in \cite{CopelandKibbleSteer,CopelandKibbleSteer2} to the study of junction dynamics in the FLRW metric.

\section{Averaged junction evolution after string collisions} \label{Averaged junction evolution after strings collision}

\subsection{Angles between strings after collision} \label{Angles between strings after collision}

Let us revisit the result~\cite{CopelandFirouzjahiKibbleSteer} that the tangent vectors $\textbf{x}^{\prime}$ of all string segments ending at a junction are coplanar in Minkowski space. We will check that this result is valid not only in Minkowski, but also in the FLRW background. The junction itself is described by the vector ${\bf X}(\tau)$ of equation (\ref{VertexEq}). Hence we have $\dot{\textbf{X}}^2 = \dot{\textbf{x}}^2_i + \dot{s}_i^2 \textbf{x}^{\prime \, 2}_i$, and together with the definition of $\varepsilon_i$ it is possible to obtain
\begin{equation}
 \begin{split}
\label{EqJunction}
 & \qquad \quad \dot{\textbf{X}}^2 = 1- \frac{\textbf{x}^{\prime \, 2}_i}{\varepsilon_i^2} + \dot{s}^2_i \textbf{x}^{\prime \, 2}_i \Rightarrow  \\
& \Rightarrow \left( \frac{1}{\varepsilon_i^2} - \dot{s}_i^2 \right) \textbf{x}^{\prime \, 2}_i = \left( \frac{1}{\varepsilon_k^2} - \dot{s}_k^2 \right) \textbf{x}^{\prime \, 2}_k.
\end{split}
\end{equation}

We now multiply equation (\ref{Boundary}) by the vector $\dot{\textbf{X}}$, sum over the index ``$i$" and use the relation (\ref{ForceBalanceEq}), to obtain
\begin{equation}
 \begin{split}
\label{Summation}
 \sum_i \varepsilon_i \dot{s}_i \mu_i = 0.
\end{split}
\end{equation}
Equation (\ref{Summation}) is just a generalized energy conservation law for shrinking and growing junctions. Using equations (\ref{Boundary}), (\ref{ForceBalanceEq}), and the definition of $\dot{\textbf{X}}$ together with the condition (\ref{Summation}), it is possible to obtain the following expression
\begin{equation}
 \begin{gathered}
\label{Triangle}
  \sum_i \varepsilon_i \mu_i \left(  \frac{1}{\varepsilon_i^2} - \dot{s}_i^2 \right) \textbf{x}^{\prime}_i = \\
    \left( \frac{1}{\varepsilon_k^2} - \dot{s}_k^2 \right) \textbf{x}^{\prime \, 2}_k \sum_i \varepsilon_i \mu_i \frac{\textbf{x}^{\prime}_i}{\textbf{x}^{\prime \, 2}_i} =0 \Rightarrow \\ 
  \Rightarrow \sum_i \varepsilon_i \mu_i \frac{\textbf{x}^{\prime}_i}{\textbf{x}^{\prime \, 2}_i} =0 .
\end{gathered}
\end{equation}

The last equation tells us that tangent vectors $\textbf{x}^{\prime }_i$ of connected strings lie in one plane, which means that three connected strings locally have to be coplanar.

\begin{figure*}
\centering
\includegraphics[width=0.46\textwidth]{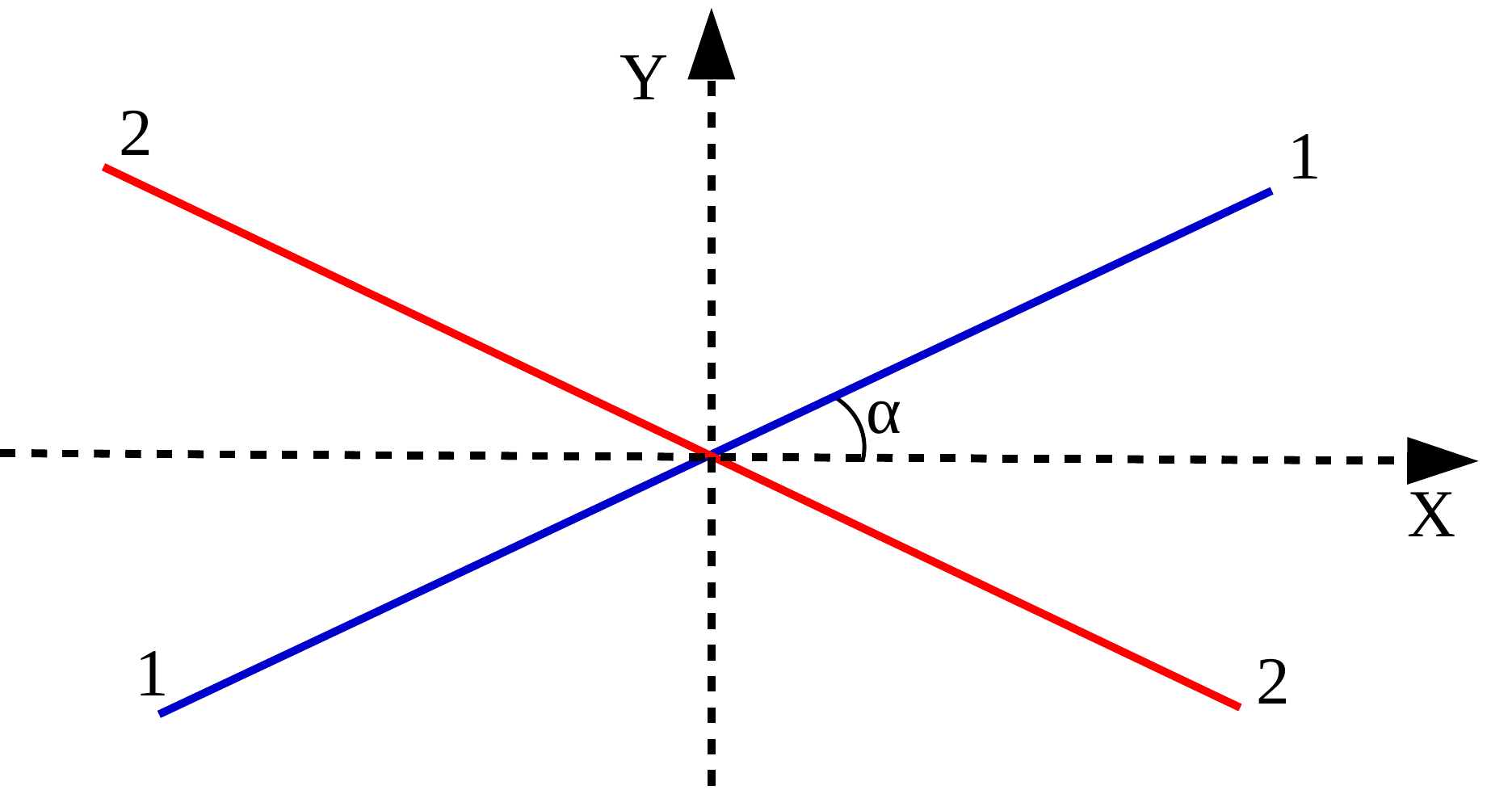}
\includegraphics[width=0.46\textwidth]{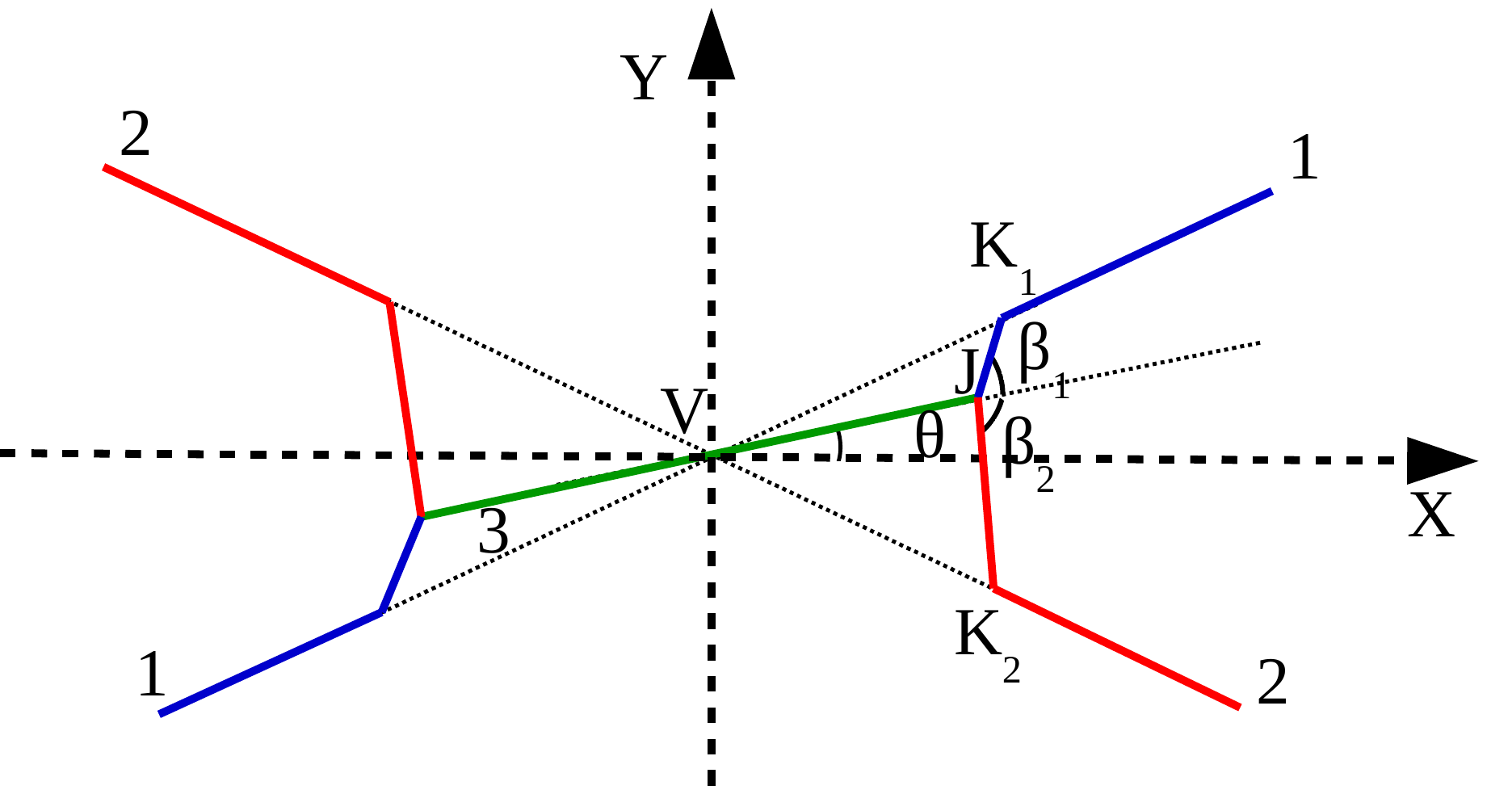}
\caption{\label{fig:Collision} Collision of two straight strings, shown as blue and red lines. When the kinematic condition $\dot{s}_3>0$ is satisfied, a pair of junctions between colliding strings is produced  with a growing string segment in-between (shown by a green line). The left panel shows the geometrical configuration of the strings just before the collision, while the right panel shows the configuration after the collision.}
\end{figure*}

To understand the string configuration right after the collision, we use the method developed in \cite{AvgoustidisPourtsidouSakellariadou}. Consider the collision of two strings with an angle $2\alpha$ between them (see figure \ref{fig:Collision}). If the kinematic condition $\dot{s}_3>0$ is satisfied, a pair of junctions, with a growing string segment in-between, should be produced at collision, and two kinks start propagating on each of the collided strings, in opposite directions, as shown in figure \ref{fig:Collision}. As a result, there will be new angles between the strings and the newly produced segment, which we denote as $\pi-\beta_1$ and $\pi-\beta_2$, with the angle between the colliding strings at the junction being $\beta_1+\beta_2$.

As was discussed above, after the collision of strings (lines $J K_1$, $J K_2$ and $V J$) stay on the same plane.
From figure \ref{fig:Collision}, following the work \cite{AvgoustidisPourtsidouSakellariadou}, we have that
\begin{equation}
\label{Betas}
\begin{gathered}
 \cos \beta_1 = \frac{V K_1 \cos (\alpha-\theta) - VJ}{J K_1} = \\
 = \frac{\cos (\alpha-\theta) - \dot{s}_3(0)}{\sqrt{1+\dot{s}_3(0) \left[ \dot{s}_3(0) - 2 \cos (\alpha-\theta) \right]}}, \\
 \cos \beta_2 = \frac{V K_2 \cos (\alpha + \theta) - VJ}{J K_2} = \\
 \frac{\cos (\alpha+\theta) - \dot{s}_3(0)}{\sqrt{1+\dot{s}_3(0) \left[ \dot{s}_3(0) - 2 \cos (\alpha+\theta) \right]}}.
\end{gathered}
\end{equation}

Hence, we obtain
\begin{equation}
\label{Betas2}
\begin{gathered}
 \tan \beta_1 = \frac{\sin (\alpha-\theta)}{\cos (\alpha-\theta) - \dot{s}_3(0)}, \\
 \tan \beta_2 = \frac{\sin (\alpha+\theta)}{\cos (\alpha+\theta) - \dot{s}_3(0)},
\end{gathered}
\end{equation}
where $\dot{s}_3(0)$ denotes the rate of junction growth at the moment of the string collision.

\begin{figure*}
\centering
\includegraphics[width=1.0\textwidth]{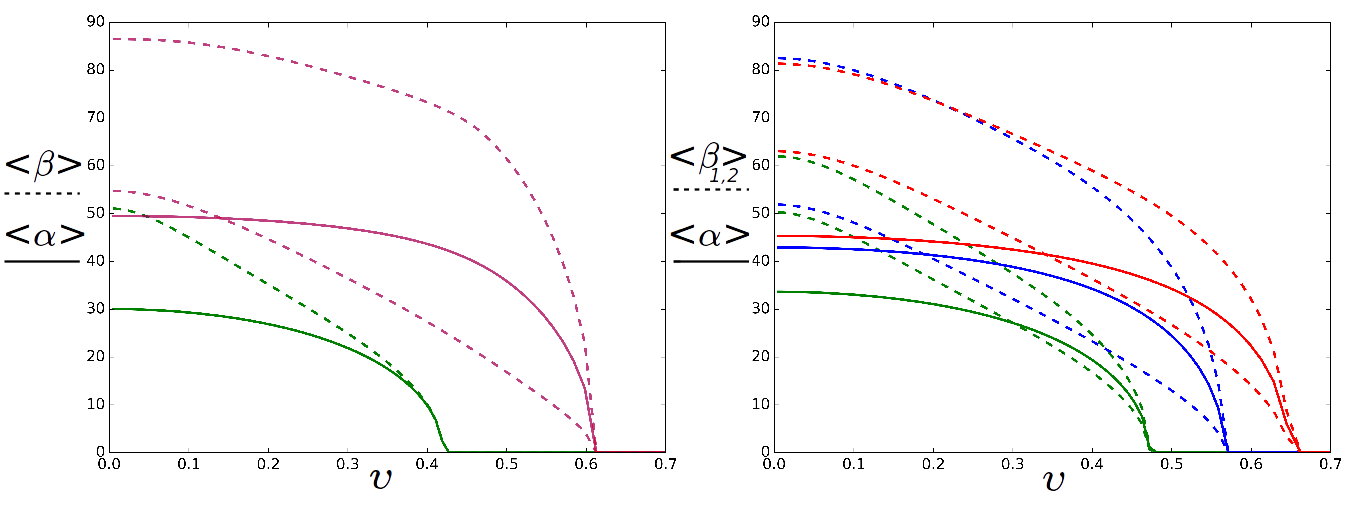}
\caption{\label{fig:Av}The collision parameters identified in figure \ref{fig:Collision} are shown here averaged over all possible angles of string collisions that give rise to junctions, cf. expression (\ref{AvProb}); the velocity distribution function has been assumed to be a Dirac delta function, and we have also chosen $P=1$ (section \ref{VOS model for strings with dynamical junctions} will provide more details about averaging). The left panel shows results for a case with two equal tensions, $\mu_1=\mu_2=1$, $\mu_3=1.4$, specifically collisions of $1,2$ (green line) and $2,3=1,3$ (purple line) pairs of strings. The right panel shows a more generic example of unequal tensions $\mu_1=1$, $\mu_2\,=\,1.2$ and $\mu_3 = 1.4$, specifically collisions of $1,2$ (green line), $1,3$ (blue line) and $2,3$ (red line) pairs of strings. Solid lines show the average angle $\alpha$ of string collisions leading to junction production. The dashed lines denote the average angles $\beta_{1,2}$.}
\end{figure*}

It should be noticed that as it was shown in \cite{AvgoustidisPourtsidouSakellariadou}, when $\upsilon \rightarrow 0$ we obtain $\beta_{1,2} = \pi/2$ for $\alpha=0$. At the same time, when $\upsilon \neq 0$, for $\alpha=0$ the angles $\beta_{1,2}=0$.

Using the equality (\ref{Betas}), it is possible to obtain the angles $\beta_{1,2}$ for all possible collisions of strings, where the full dependence on the tensions and relative velocities of the colliding strings becomes apparent through equation (\ref{s3}). Since we are interested in the average picture over a string network, we wish to know how angles $\beta_{1,2}$ depend on the rms string velocity for fixed values of the string tensions $\mu_1$, $\mu_2$ and $\mu_3$. In order to achieve this goal we need to express the angle $\theta$ and junction growth rate $\dot{s}_3$ as functions of $\alpha$, $\upsilon$ and $\mu_i$ ($i=1,2,3$). For this purpose we use the following equations for straight strings \cite{CopelandKibbleSteer}
\begin{equation}
\label{JunctEqs}
\begin{gathered}
 \left[ \mathcal{M} \dot{s}_3 + M_3 (1-c_3)  \right] \gamma_u^{-1} \cos \theta = \\
 = \left[ M_1 (1-c_1)+M_2 (1-c_2) \right] \gamma_{\upsilon}^{-1} \cos \alpha, \\
 \left[ \mathcal{M} \dot{s}_3 + M_3 (1-c_3)  \right] \gamma_u^{-1} \sin \theta = \\
 = \left[ M_1 (1-c_1)-M_2 (1-c_2) \right] \gamma_{\upsilon}^{-1} \sin \alpha, \\
\left[ \mathcal{M} - M_3 (1-c_3)  \right] u = \\
 =\left[ M_1 (1-c_1)-M_2 (1-c_2) \right] \upsilon
\end{gathered}
\end{equation}
from which we can also obtain
\begin{equation}
\label{JunctEqs2}
\begin{gathered}
   \theta = \arctan \left( \frac{u}{\upsilon} \tan \alpha \right), \\
 \left[\mu_3^2 (1-\upsilon^2)+\mu_{-}^2(\upsilon^2 \cos^2 \alpha - \sin^2 \alpha)  \right] u^2  +\\
 + \mu_{-} \sin^2 \alpha u^4 - \mu_{-}^2 \upsilon^2 \cos^2 \alpha=0. 
\end{gathered}
\end{equation}

Hence, we can numerically obtain the angles $\beta_{1,2}$ as functions of $\mu_i$, $\upsilon$ and $\alpha$. Taking the integral over the angles $\alpha$ from $0$ to the maximal value of $\alpha_{cr}$ (the maximum angle at which a junction can be created) we average over all possible collisions that lead to junction production (see section \ref{VOS model for strings with dynamical junctions} and reference  \cite{PourtsidouAvgoustidisCopelandPogosianSteer} for more details). 

It should be noted that the velocity $\upsilon$ used up to now (see equation (\ref{ThreeStrings}) describing two colliding straight string segments) is the velocity in the frame where both strings move towards each other with equal speeds $\upsilon$. However, when we consider the encounters of a typical string with other strings in the network (integrating over all possible relative angles and velocities) we need to transform our calculations to the rest frame of the string in which the collision velocity $v_{cl}$ is related to $v$ by
\begin{equation}
\label{VelTransform}
v_{cl} = \frac{2 \upsilon}{1+\upsilon^2} \,.
\end{equation}

Further in the text, individual string velocities will be defined in the string rest frame as collision velocities but denoted simply as $v$. See appendix \ref{APDX} for for further clarification of the different velocities used in this work.

For now the velocity distribution is taken in the form of a delta function centred in the rms velocity value (but can be generalised to a gaussian centred at the rms string velocity with a width introduced as an additional parameter).

Carrying out the numerical computation of the average values of the angles $\beta_{1,2}$, we build the plot of the angle-dependence on the rms velocity. In figure \ref{fig:Av} we show examples of the average value of $\beta$ (for $\mu_1=\mu_2$) and $\beta_1$, $\beta_2$ (for $\mu_1 \neq \mu_2$) depending on the rms velocity $\upsilon$ of the string network. The results for the average angles $\beta_1$ and $\beta_2$ will be used later as initial values for the study of the subsequent dynamics of junctions in the string network.

\subsection{Correlation functions of ingoing components and small-scale structure} \label{Correlation function of incoming components; small-scale structure}

To understand junction growth in the string network, we need to figure out the behaviour of the averaged scalar product that occurs from averaging equations (\ref{JunctionsUnit3}) and (\ref{c_i})
\begin{equation}
\begin{split}
\label{h}
h_i = 1 - < c_i >\,,
\end{split}
\end{equation}
where $c_i$ is defined in Eq. (\ref{c_i}). In other words we need to model how these scalar products evolve and study their effect on junction growth through equations (\ref{JunctionsUnit3})

\begin{equation}
\begin{split}
\label{JunctionDynAv}
 & \left\langle \varepsilon_i \dot{s}_i \mu_i \mathcal{M} \right\rangle =  \mu_i \mathcal{M}_h  - \mu M_i h_i,
\end{split}
\end{equation}
where
\begin{equation}
\label{AverDef}
 \left\langle ... \right\rangle = \frac{\int ... \, \varepsilon d \sigma }{\int \varepsilon d \sigma}
\end{equation}
denotes the average of the enclosed quantity over the string network and
\begin{equation}
\begin{gathered}
\label{AverDefM}
  \mathcal{M}_h = \frac{\int \varepsilon_1 \varepsilon_2 \varepsilon_3  \sum_j M_j (1 - c_j) \, d \sigma_1 d \sigma_2 d \sigma_3 }{\int \varepsilon_1 d \sigma_1 \int \varepsilon_2 d \sigma_2 \int \varepsilon_3 d \sigma_3} = \\
= \sum_j M_j h_j.
\end{gathered}
\end{equation}

Here, the averages have been taken over corresponding string types and we assumed that $c_i$ depends only on $\sigma_i$ (correlation function changes with junction length growth; see \ref{Junction rates for a scaling string network} for more details).

We should emphasize that the correlator $h$ is defined between different strings connected at the junction: $h$ includes product of vectors $\textbf{p}_i$ that belong to different strings. Below we will also use the correlator $\text{h}$ defined on one string, which includes the product of vectors $\textbf{p}_i$ that belong to one string separated by some distance $l$.

To approach this problem, we can study the dependence of the scalar product $<\textbf{p}(x_1)_i \cdot \textbf{p}(x_2)_i>$ on the distance $l=x_2-x_1$. This problem has been studied both numerically and analytically in a number of papers \cite{KibbleCopeland, AllenCaldwell,  VincentHindmarshSakellariadou, MartinsShellard_Fractal,PolchinskiRocha, KibbleCopeland2, VanchurinOlumVilenkin, HindmarshStuckeyBevis}. We revisit the main results of these studies and emphasize the important points for further investigation.

Let us recall the definition  of $\varepsilon$ and notice that this variable depends on the $\sigma$ parametrization. Thus, if the length of the string is reduced by a factor $p=\sigma/\tilde{\sigma}$, we have the following relation
\begin{equation}
\label{Preduced}
\varepsilon(\tilde{\sigma}) = \sqrt{ \frac{\partial_{\tilde{\sigma}}  \textbf{x}^2 }{1 - \dot{\textbf{x}}^2}} = \frac{\partial \sigma}{\partial \tilde{\sigma}} \sqrt{ \frac{\partial_{\sigma}  \textbf{x}^2 }{1 - \dot{\textbf{x}}^2}} = p \, \varepsilon(\sigma).
\end{equation}

We see that the equation of motion (\ref{MicroEqOfMotString}) is invariant under the transformation (\ref{Preduced}) if the multiplier $p$ is a constant. However, if $p$ is a time dependent function, there will be an additional contribution in the form
\begin{equation}
\label{EnergMicroP}
 \dot{\varepsilon}_i+2 \frac{\dot{a}}{a} \varepsilon_i \dot{\textbf{x}}_i^2 + \frac{\dot{p}}{p}\varepsilon_i= 0. \\
\end{equation}

To estimate the ratio $\dot{p}/p$ we should remember that it is proportional to the string energy, i.e. the larger $p$ is, the more energy the string has. Using the effective chopping parameter describing the average probability for one string to lose energy due to loop production, we can write 
\begin{equation}
\label{Pestimation}
\frac{\dot{p}}{p} = \frac{\dot{\rho}}{\rho} \bigg\vert_{\text{loops}} = c_l \frac{\vert \upsilon \vert}{L},
\end{equation}
where in the last equality we used the standard contribution to energy loss from the loop production mechanism, where the length $L_{c \, i}$ is the comoving distance that the string travels before a collision (average comoving distance between strings). This is akin to the mean free path. Note, however, that this is a small-scale-structure energy loss mechanism, that can be understood as reducing the string energy per correlation length (without affecting the string number density) rather than directly altering the correlation length. 

Using the relation (\ref{Pestimation}), we can include the phenomenological energy loss term at the microscopic level. Equation (\ref{EnergMicroP}) can then be rewritten as 
\begin{equation}
\label{EnergMicroEq}
 \dot{\varepsilon}_i+2 \frac{\dot{a}}{a} \varepsilon_i \dot{\textbf{x}}_i^2 + \varepsilon_i \vert \upsilon_i \vert \frac{c_{l 
 \, i}}{L_{c \, i}}= 0. \\
\end{equation}

It can be shown that using equation (\ref{EnergMicroEq}), one can rewrite the equations of motion for a string in terms of vectors $\textbf{p}$ and $\textbf{q}$ \cite{AlbrechtTurok} as
\begin{equation}
\label{EquationOfMotPQ}
\begin{gathered}
 \dot{\textbf{p}}_{i}+\frac{1}{\varepsilon_{i}} \textbf{p}^{\prime}_{i} = \frac{\dot{a}}{a} \left( \textbf{q}_{i} - (\textbf{p}_{i} \cdot \textbf{q}_{i}) \textbf{p}_{i} \right) + \\
 + \frac{\textbf{q}_{i}+\textbf{p}_{i}}{2}\sqrt{\frac{1-(\textbf{p}_{i} \cdot \textbf{q}_{i})}{2}} \frac{c_{l \, i}}{L_{i}}, \\
 \dot{\textbf{q}}_{i}-\frac{1}{\varepsilon_{i}} \textbf{q}^{\prime}_{i} = \frac{\dot{a}}{a} \left( \textbf{p}_{i} - (\textbf{p}_{i} \cdot \textbf{q}_{i}) \textbf{q}_{i} \right) + \\
 + \frac{\textbf{q}_{i}+\textbf{p}_{i}}{2}\sqrt{\frac{1-(\textbf{p}_{i} \cdot \textbf{q}_{i})}{2}} \frac{c_{l \, i}}{L_{i}}.
\end{gathered}
\end{equation}

\begin{figure}
\centering
\includegraphics[width=0.46\textwidth]{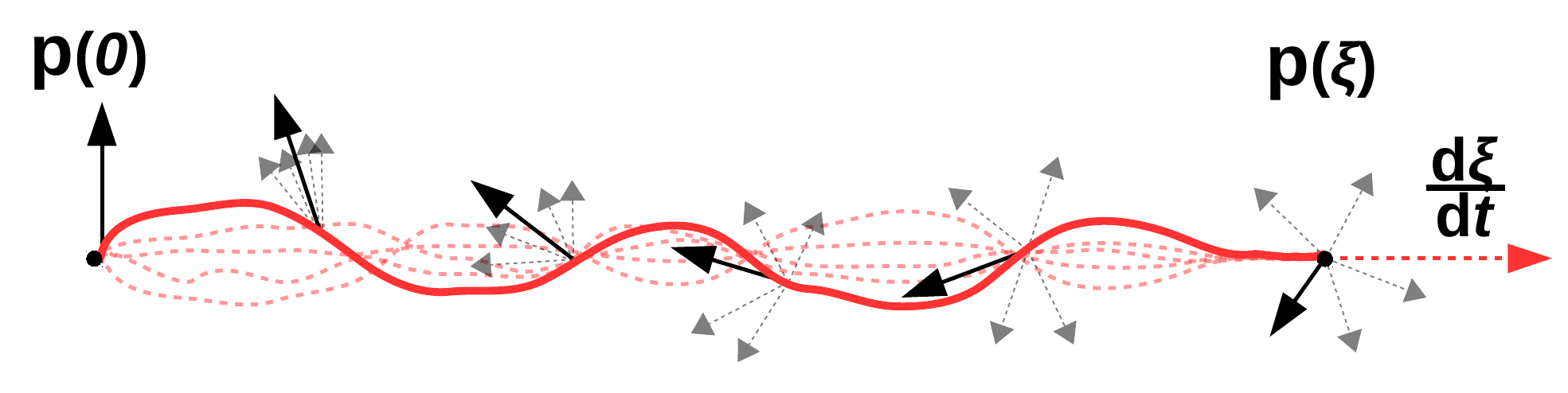}
\caption{\label{fig:pEvol} Schematic evolution of the vector $\textbf{p}$ along the string. Dashed lines show different possible realizations of the string with corresponding faded vectors $\textbf{p}$. At the correlation length $\xi$ vectors $\textbf{p}$ become, on average, completely independent $< \textbf{p}(0) \cdot \textbf{p}(\xi)> = 0$. Note that the correlation length $\xi$ is a function of time, which grows in an expanding universe.}
\end{figure}

In the same way as it was done in \cite{PolchinskiRocha}, we can choose the characteristic variable $s_{\sigma}(\tau)$ instead of $\sigma$, which is constant for ingoing waves ($\partial_{s_{\sigma}} \textbf{p} = 0$). Using this new variable $s_{\sigma}(\tau)$, let us estimate how the orientation of the vector $\textbf{p}_i$ is changing along the string (see figure \ref{fig:pEvol}). In particular, we are interested in the average scalar product (correlation) between two vectors $\textbf{p}_i$ as a function of their separation and time, $\text{h}_i( l ,t)$,
\begin{equation}
\text{h}_i = 1 - <\textbf{p}_i \cdot \tilde{\textbf{p}}_i> \,.
\end{equation}
Note the roman font for this correlator (${\rm h}_i$), indicating that both vectors are on the same string segment. This is in contrast to the correlator $h_i$ (italic font) encountered previously (cf. \ref{h}), in which the two vectors belong to two different strings.

For this purpose we are going to choose the $\textbf{p}_i$ parametrizations for the two vectors as $s_{\sigma} (\tau)$ and $s_{\sigma}'(\tau)$, separated by the physical length $l$. It should be noticed that the time dependence of parameter $s_{\sigma} (\tau)$ does not change equations (\ref{EquationOfMotPQ}). In order to understand how the average scalar product is evolving with time, and how it depends on length, we write the averaged equation for the time evolution of the scalar product (further we will write $s_{\sigma}$ instead of $s_{\sigma}(\tau)$ for compactness):

\begin{equation}
\begin{gathered}
\label{VectEvol}
  < \partial_{\tau} \left( \textbf{p}_i \cdot \tilde{\textbf{p}}_i \right)>   = \frac{ <\textbf{q}_i \cdot \tilde{\textbf{p}}_i + \textbf{p}_i \cdot \tilde{\textbf{p}}_i > }{2} \vert \upsilon_i \vert \frac{c_{l \, i}}{L_i}+ \\
 \frac{ < \tilde{\textbf{q}}_i \cdot \textbf{p}_i + \textbf{p}_i \cdot \tilde{\textbf{p}}_i > }{2} \vert \upsilon_i \vert \frac{c_{l \, i}}{L_i}  +\\
  \frac{\dot{a}}{a} < \textbf{q}_i \cdot \tilde{\textbf{p}}_i + \textbf{p}_i  \cdot \tilde{\textbf{q}}_i  -  2 \textbf{p}_i \cdot  \tilde{\textbf{p}}_i   \alpha_i > + \\
  + \sum_j^{N_k} < \delta(s-s_j) \textbf{p}_i \cdot \tilde{\textbf{p}}_i  + \delta(s'-s'_j) \tilde{\textbf{p}}_i \cdot \textbf{p}_i > ,
\end{gathered}
\end{equation}
where $\alpha_i = <\textbf{p}_i \cdot \textbf{q}_i> =1 - 2 \upsilon_i^2 $ with $\upsilon_i^2 = <\dot{\textbf{x}}_i^2 >$, tilded and untilded variables are calculated at $s'$ and $s$ respectively, the sum appears due to the possible presence of kinks on the length between $s$ and $s'$, $N_k$ is the number of kinks between $s$ and $s'$, $\delta(...)$ is the Dirac delta function, and the sign $< ... >$ means averaging over an ensemble of segments and integrating over many Hubble times.

To treat equation (\ref{VectEvol}) we are going to use the method employed in \cite{PolchinskiRocha}. Let us consider that $s_{\sigma}$ and $s_{\sigma}'$ are close to each other, and we expand equations (\ref{VectEvol}) around $s_{\sigma} - s_{\sigma}'$ and drop all terms higher than first order in $[s_{\sigma}-s_{\sigma}']$. The outgoing $\textbf{q}(s_{\sigma},\tau)$ and ingoing $\tilde{\textbf{p}}(s_{\sigma}',\tau)$ modes meet each other at the worldsheet point $(s_{\sigma},\tau-\delta)$, where $\delta$ is of the order of $[s_{\sigma}-s_{\sigma}']$. Hence, the product of outgoing and ingoing modes can be approximated as $< \textbf{q}_i(s_{\sigma},\tau) \cdot \tilde{\textbf{p}}_i(s_{\sigma}',\tau) > = \alpha_i $ (for more details see \cite{PolchinskiRocha}).

Moreover, the averaged contribution from the sum of kinks is proportional to the linear density of kinks $K(\tau)$, which also includes the average sharpness of kinks, multiplied by the interval $(s-s')$.  As a result, equation (\ref{VectEvol}) can be rewritten as 
\begin{equation}
\begin{gathered}
\label{VectEvol2}
 \partial_{\tau} \text{h}_i  = - \text{h}_i  \left( 2 \frac{\dot{a}}{a} \alpha_i - \frac{\vert \upsilon_i \vert c_{l\, i}}{L_i} \right) - \\
 - \frac{\vert \upsilon_i \vert c_{l \, i}}{L_i} \left( 1 + \alpha_i \right) + 2  K (1-\text{h}_i) + O([s_{\sigma}-s_{\sigma}']^2)\,.
\end{gathered}
\end{equation}
We emphasize that the quantitative difference between ${\rm h}_i$ (relevant here) and ${h}_i$ (cf. Eq. \ref{h}) is that when the distance $l$ approaches zero the former necessarily approaches zero (since the correlator approaches unity), while the latter need not do so since the value of the correlator near the junction is determined by the string junction configuration described in Sect. \ref{Angles between strings after collision} (note that the product of vectors $\textbf{p}_i$ that belong to different strings at the junction does not have to be unity).

Assuming that the scale factor evolves according to a power law $a \propto \tau^n$, we anticipate that the characteristic length $L_i \propto \epsilon_i \tau$ and the kink density is $K(t) \propto \frac{k_k}{\tau}$ (the parameter $k_k$ denotes the kink decay rate, which can be caused by expansion, radiation \cite{ AllenCaldwell} or backreaction -- this requires further study to understand how fast kinks can be smoothed; recent results on this subject can be found in \cite{Blanco-PilladoOlumWachter, ChernoffFlanaganWardell}) where $n$, $\epsilon_i$ and $k_k$ are constants. With these assumptions we can solve (\ref{VectEvol2}) finding 
\begin{equation}
\label{CorrelaSol}
\text{h}_i \propto \frac{g_i (s_{\sigma}' - s_{\sigma})}{\tau^{2 (n \overline{\alpha}_i + k_k) - \vert \upsilon_i \vert c_{l \, i} / \epsilon_i }}\,,
\end{equation}
where $g_i (s_{\sigma}' - s_{\sigma})$ is a function that depends only on the distance along the string, defined by parameters $s_{\sigma}' - s_{\sigma}$.

At the same time, under our assumptions, we can establish from equation (\ref{EnergMicroEq}) that
\begin{equation}
\label{EpsSol}
\varepsilon_i \propto \tau^{-2 n \upsilon_i^2 + \vert \upsilon_i \vert c_{l \, i} / \epsilon_i}
\end{equation}
and so the corresponding string length has the form
\begin{equation}
\label{StrLength}
l = a \int \epsilon d \sigma \propto a (\sigma-\sigma') \tau^{-2 n \upsilon_i^2 + \vert \upsilon_i \vert c_{l \, i} / \epsilon_i}.
\end{equation}

Assuming that the length $l$ has a scaling behavior $l \propto x_0 t$, where $x_0$ is a constant, we obtain the relation
\begin{equation}
\label{StrLengthSigma}
\sigma' - \sigma = x_0 \tau^{1 + 2 n \upsilon_i^2 - \vert \upsilon_i \vert c_{l \, i} / \epsilon_i}.
\end{equation}

Using the property that $\sigma-\sigma' = s-s'$ and substituting (\ref{StrLengthSigma}) in the solution (\ref{CorrelaSol}) as initial condition for a specific constant value $\text{h}_i = \text{h}_0$, we obtain that \cite{PolchinskiRocha}
\begin{equation}
\label{CorrelaSol2}
\text{h}_i = \text{h}_0 \left( \frac{\sigma - \sigma'}{x_0 \tau^{1+2 n \upsilon_i^2 - \vert \upsilon_i \vert c_{l \, i} / \epsilon_i}} \right)^{2 \chi_i} = A  \left( \frac{l}{t} \right)^{2 \chi_i},
\end{equation}
where $\chi_i = \frac{ n \overline{\alpha}_i + k_k - \kappa_i}{1+2 n \upsilon_i^2 -2 \kappa_i} $ with $\kappa_i = \frac{\vert \upsilon_i \vert \tilde{c}_i}{2 \epsilon_i}$, and $A$ is a constant.

\begin{figure*}
\centering
\includegraphics[width=0.47\textwidth]{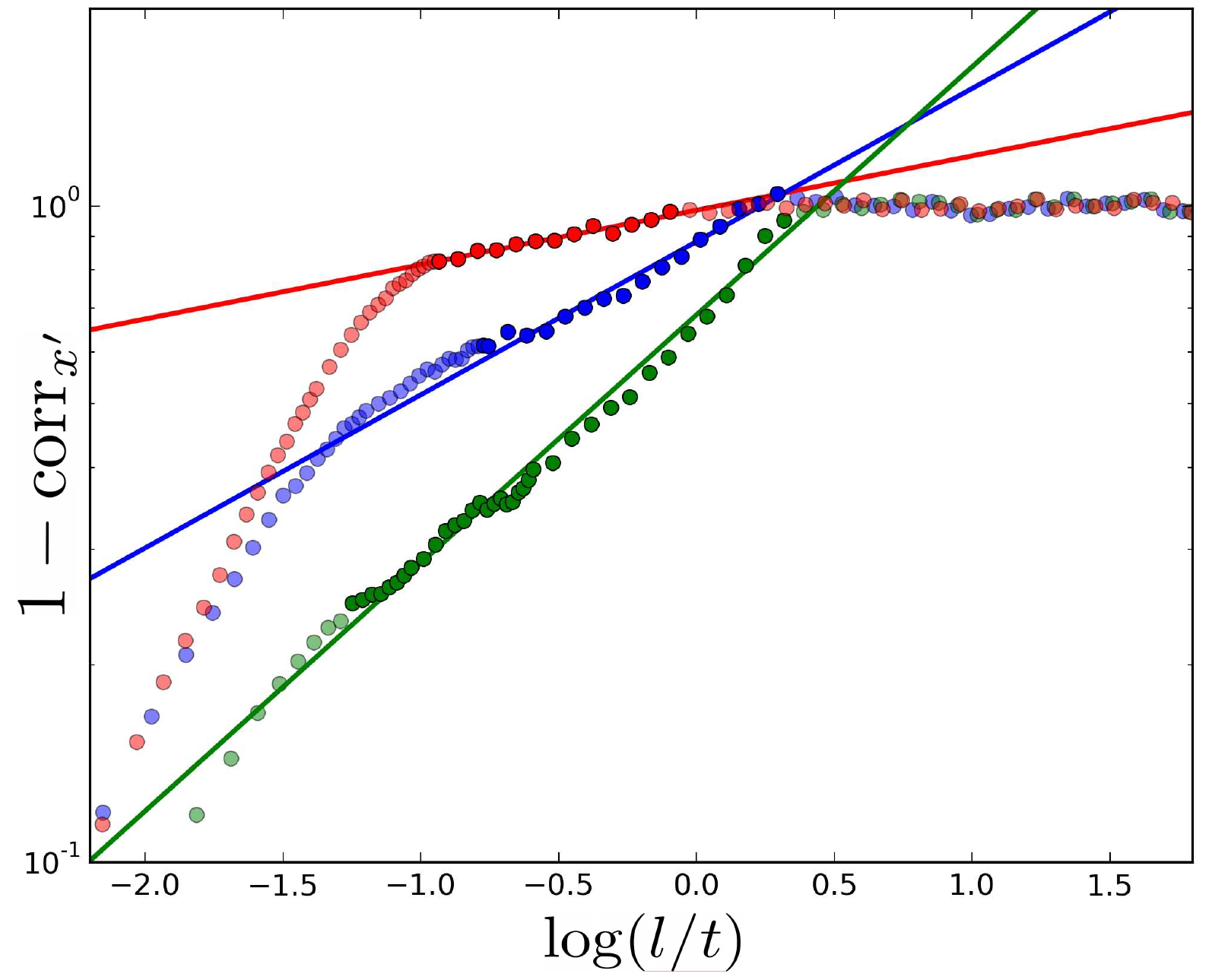}
\includegraphics[width=0.47\textwidth]{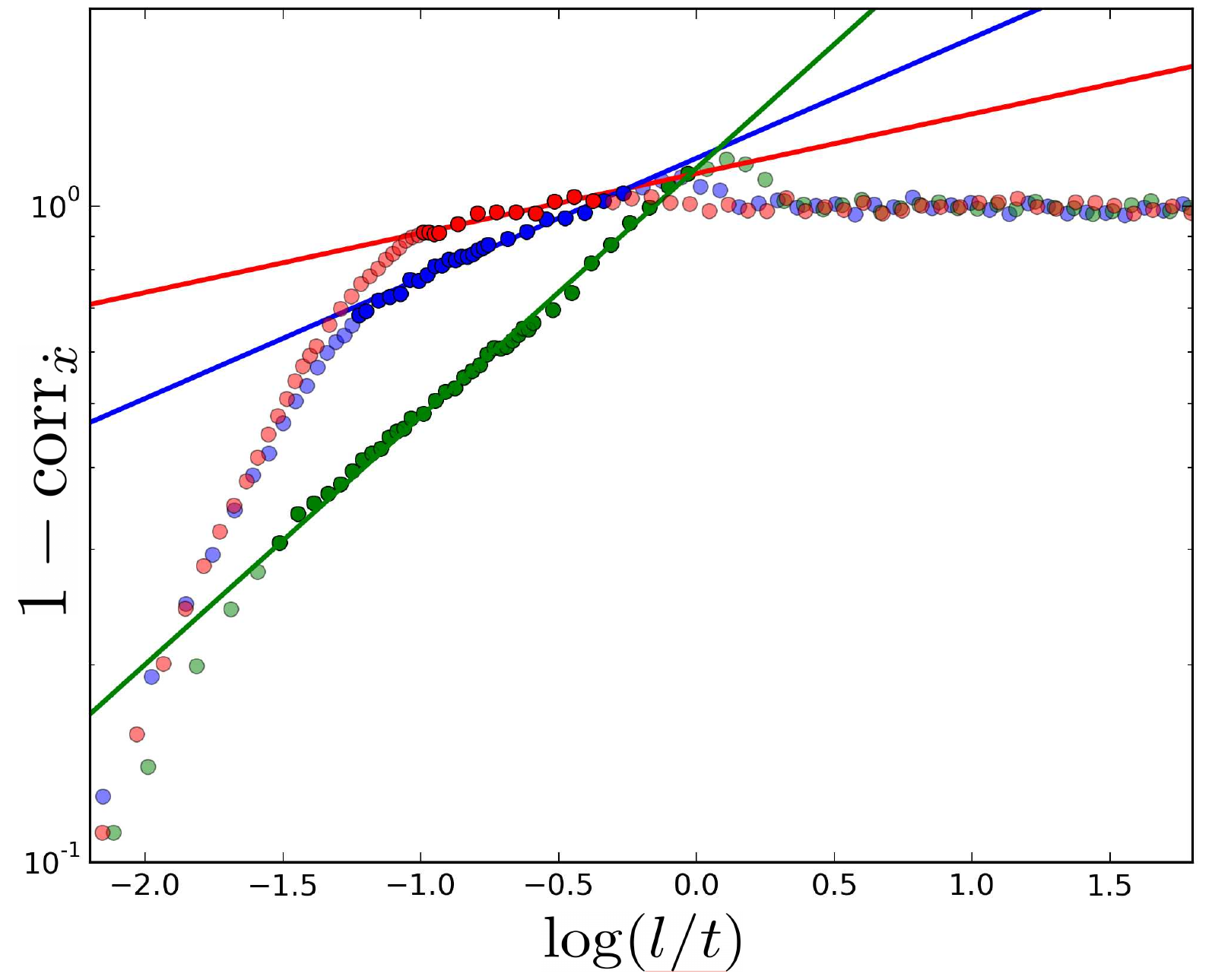}
\caption{\label{fig:CorrX-V}The behaviour of tangent and velocity correlators. The shaded dots represent data from simulations \cite{MartinsShellard_Fractal} for the correlation functions (\ref{CorrX}) and (\ref{CorrV}), depending on the string length $l$ and physical time $t$. To fit the analytic estimate for these correlation functions we use the interval of $l/t$ that has already reached scaling behaviour (\ref{CorrelaSol2}). The data points used for the fits dots are shown by dots in bright color. Solid lines demonstrate the best fit of the correlation functions (\ref{CorrX}), (\ref{CorrV}) for parameters $A$ and $\chi$. Different colors correspond to different expansion rates: red $n=0$, blue $n=1$, green $n=2$.}
\end{figure*}

Comparing (\ref{CorrelaSol2}) with the corresponding result from \cite{PolchinskiRocha}, we see that due to including the loop production term and kinks, we have additional terms in $\chi_i$. These are essential for describing the evolution of the correlation function (\ref{CorrelaSol2}) in Minkowski space. We note that the result obtained in \cite{PolchinskiRocha} would not apply to the case of Minkowski space (for which it would give the clearly incorrect $\chi_i=0$) while our result is also valid for that case.

Let us compare the equation (\ref{CorrelaSol2}) with the results of Nambu-Goto simulations \cite{MartinsShellard_Fractal}. To do so, we are going to use simulations for string network evolution for all currently available expansion rates: $n=0$ (Minkowski space), $n=1$ (Radiation domination epoch), $n=2$ (Matter domination epoch). We fix the values of the chopping parameters $c_l$ for each expansion rate by studying the corresponding scaling regimes, in accordance with \cite{MartinsShellard, MartinsShellard2}. Taking fixed values of $c_l$ we are left with only one additional free parameter $k_k$ in equation (\ref{CorrelaSol2}). To obtain $k_k$ we are going to fit a dependence of the form (\ref{CorrelaSol2}) to simulations measurements of the correlators of the tangent vectors \cite{PolchinskiRocha}
\begin{equation}
\label{CorrX}
\begin{gathered}
\text{corr}_{x^{\prime}} = \frac{<\textbf{x}^{\prime}(\sigma,\tau) \cdot \textbf{x}^{\prime}(\sigma',\tau)>}{<\textbf{x}^{\prime}(\sigma,\tau) \cdot \textbf{x}^{\prime}( \sigma,\tau)>} \approx \\
\approx 1 - \frac{A}{2(1-\upsilon^2)} \left( \frac{l}{t} \right)^{2 \chi }
\end{gathered}
\end{equation}
and velocities along the string
\begin{equation}
\label{CorrV}
\begin{gathered}
\text{corr}_{\dot{x}} = \frac{<\dot{\textbf{x}}(\sigma,\tau) \cdot \dot{\textbf{x}}(\sigma',\tau)>}{<\dot{\textbf{x}}(\sigma,\tau) \cdot \dot{\textbf{x}}( \sigma,\tau)>} \approx \\
\approx 1 - \frac{A}{2 \upsilon^2} \left( \frac{l}{t} \right)^{2 \chi }.
\end{gathered}
\end{equation}
The results of this analysis are depicted in figure \ref{fig:CorrX-V} and the values obtained for the fitted model parameters are listed in table \ref{TableMeasureChiA}. 

\begin{table}
\centering
\caption{Fitted parameters of the correlation function (\ref{CorrelaSol2}) for different expansion rates ($n=0,1,2$), from independent measurements of correlators $\text{corr}_{x^{\prime}}$ (Eq. \ref{CorrX}) and $\text{corr}_{\dot{x}}$ (Eq. \ref{CorrV}) (with corresponded indices) obtained from Nambu-Goto string simulations \cite{MartinsShellard_Fractal}. The loop chopping parameters are fitted from the scaling regimes of the networks using the VOS equations for $L$. }
\label{TableMeasureChiA}
  \begin{tabular}{ l | c | c | c | c | c | c | c }
    \hline
     & $A_{x^{\prime}}$  & $A_{\dot{x}}$ & $\chi_{x^{\prime}}$ & $\chi_{\dot{x}}$ & $k_{k \, x^{\prime}}$ & $k_{k \, \dot{x}}$ & $c_l$ \\ \hline
    $n=0$ & $1.37$ & $1.16$  & $0.04$ & $0.05$ &  $0.96$ & $0.95$ & $0.33$ \\ \hline
    $n=1$ & $1.39$ & $1.22$ & $0.12$ & $0.09$ & $0.48$ & $0.47$ & $0.27$  \\ \hline
    $n=2$ & $1.03$ & $0.92$ & $0.20$ & $0.19$ & $0.04$ & $0.03$ & $0.24$  \\ \hline
  \end{tabular}
\end{table}

A similar scaling behaviour for $l/t$ has been observed in numerical simulations of Abelian Higgs cosmic strings \cite{HindmarshStuckeyBevis}. The typical values of the parameters $\chi_i$ measured in Abelian-Higgs simulations are higher than in Nambu-Goto. While the most parsimonious explanation for this difference stems from the fact that Abelian-Higgs simulations have a much lower spatial resolution (making the measurement of these correlators harder and less accurate), it  could also be explained by domination of another type of energy loss mechanism or by a difference in the evolution of kinks. 

It is worthy of note that the parameter $\chi$ appears in a number of relevant string network characteristics, notably the fractal dimension and the loop distribution function: for an approximate analytic description see \cite{PolchinskiRocha, Rocha, DubathPolchinskiRocha} and for numerical simulation results see \cite{MartinsShellard_Fractal, RingevalSakellariadouBouchet, HindmarshStuckeyBevis, Blanco-PilladoOlumShlaer}. As a result, the understanding and accurate modelling of parameter $\chi$ is important for quantifying observational outcomes of cosmic string networks \cite{RingevalSuyama}.  

An important remark on figure \ref{fig:CorrX-V} is that the correlation function $\text{h}$ reaches unity on the length $l \approx t$. This fact is explicitly seen for Nambu-Goto and Abelian-Higgs simulations if the correlators are plotted against $\log (l/t)$ instead of the ratio between $l$ and characteristic lengthscale $L$. Hereinafter, we will use this as our definition of the correlation length $\xi$, i.e. the distance $l$ at which the vectors $\textbf{x}^{\prime}$ and $\dot{\textbf{x}}$ become uncorrelated. (Note that there are several different definition in the literature.) Thus, with our definition, simulations show that $\xi \approx t$.

We should remember that the treatment described above is an approximation, valid within a specific range of scales. If we consider the smallest scales, where loop production is insignificant and only one kink is present, then, as was shown in \cite{KibbleCopeland2}, the correlator $\text{h}_i$ has the following behaviour
\begin{equation}
\label{CorrelaSolCopKib}
\text{h}_{i \, kink} \propto \left( \frac{l}{t} \right).
\end{equation}
The relation (\ref{CorrelaSolCopKib}) can be evolving  slightly during network evolution. However, we do not anticipate significant corrections when the network reaches a scaling regime.

\begin{figure}
\centering
\includegraphics[width=0.49\textwidth]{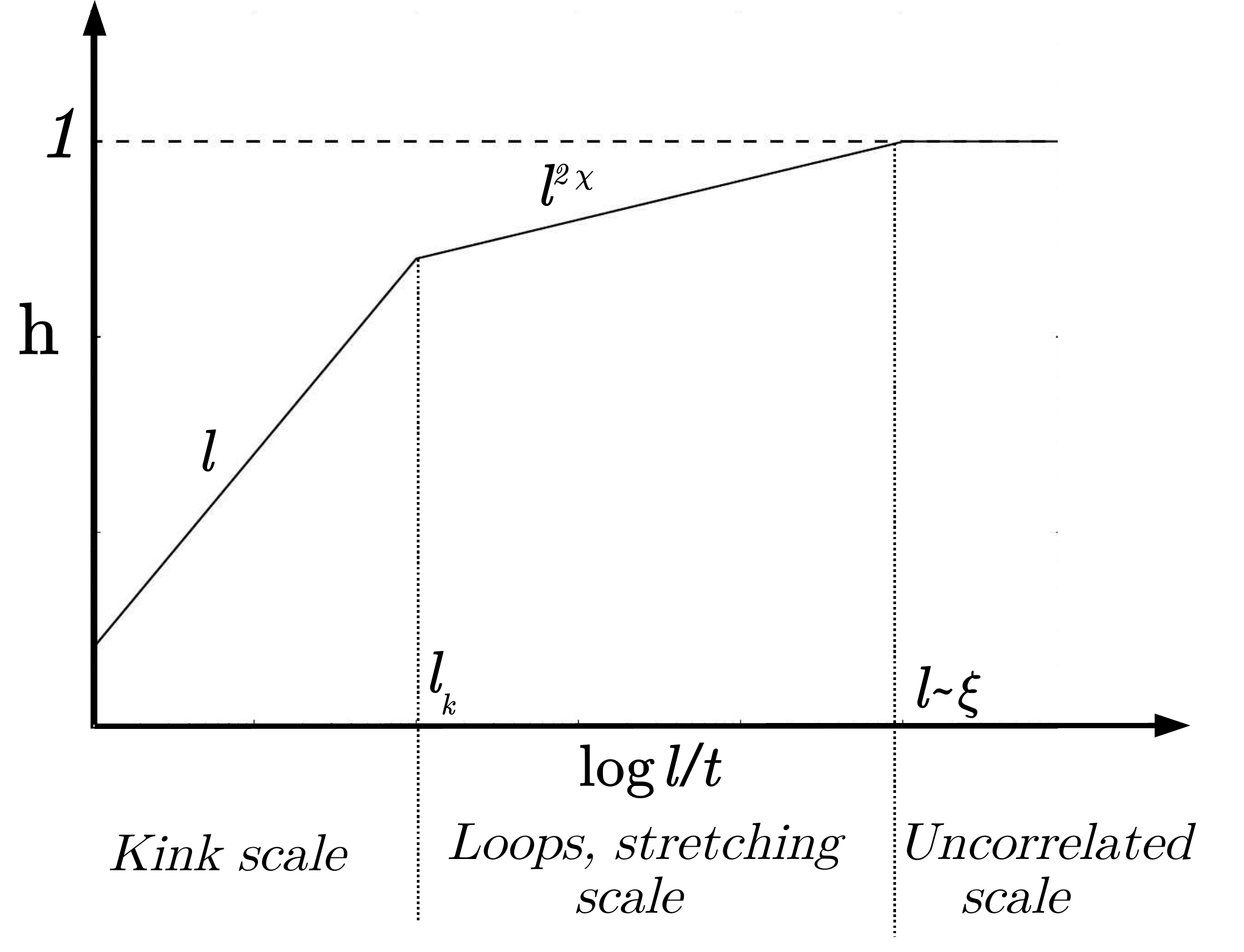}
\caption{\label{fig:Corr}  Behaviour of the function $\text{h}(l/t)$. The plot shows how the correlation between vectors decreases with distance $l$ (logarithmic scale). On small sales the function should be linear $\text{h} \propto l/t$, but then it becomes a power law $\text{h} \propto (l/t)^{2 \chi}$, with $\chi$ a described in the text. Eventually the function $\text{h}$ becomes constant $1$ when the distance $l \sim \xi$, the correlation length. }
\end{figure}

For scales larger than the correlation length, $l/\xi>1$, the function $\text{h}_i$ tends to unity (since the correlator itself vanishes). We are interested in finding a function that can mimic all ranges of $l/t$ for the correlation function $\text{h}_i$. We denote this by $\text{h}(l / t)$ and we sketch its behaviour in figure \ref{fig:Corr}.

\section{Extended VOS model with dynamical junctions} \label{VOS model evolution with dynamcial junctions}

\subsection{Junction increase/decrease rates for a scaling string network} \label{Junction rates for a scaling string network}

In the previous section we estimated the behaviour of the correlation function $\text{h}(l/t)$ on a string. Now, using this result, we are ready to study  junction evolution when the string network reaches a scaling regime, i.e. when the characteristic length and rms velocity behave as $L \propto \epsilon \tau$, $\upsilon = \text{const}$. To do so, we use the average equation (\ref{JunctionDynAv}), where the functions $h_i$ are approximated by the function $\text{h}_i$ depicted in figure \ref{fig:Corr} (recall that the only difference between the two is that in the limit of zero separation only the latter needs to vanish). 

We define the average rate of change of the string conformal length as (compare with the definition (\ref{StrLength}) for the string length)
\begin{equation}
\begin{split}
\label{LengthChange}
 & \Delta \dot{l}_{c} = \left\langle  \varepsilon \dot{s} \right\rangle.
\end{split}
\end{equation}
In an averaged sense, we expect a decay of correlation when the junction destroys or produces strings. The way this happens should be (again, in a statistical sense) similar to the way the correlation function decays along the string, when we compare different segments of it. Since the new strings (produced due to collisions) should initially grow ($\dot{s}_i>0$), it seems reasonable to connect the length of this string growth with the decay of correspondent correlation.

We thus make the further simplifying but reasonable assumption that the correlation function $h_i$ is determined by the change of comoving length $\Delta l_{c \, g}$ of the growing string (the one that is being created by string zipping process). Using definitions (\ref{AverDef}),(\ref{LengthChange}), we can then average equations (\ref{JunctionDynAv}) to find
\begin{equation}
\begin{split}
\label{LengthChangeEq}
 & \Delta \dot{l}_{c \, i}^{(n)} = 1 - \frac{ \mu M_i h_i( \Delta l_{c \, g}^{(n)} / \tau)}{\mu_i \mathcal{M}},
\end{split}
\end{equation}
where the indices $n$ count the number of possible string collisions and the subscript $g$ indicates the newly created string that is growing after junction formation while the colliding strings are zipping along their length. To obtain (\ref{LengthChangeEq}) we also assumed that
 $< \varepsilon_i \dot{s}_i (1-c_i) > = \Delta \dot{l}_{c \, i} h_i$

We should keep in mind that we are interested in the behaviour of the correlation function in figure \ref{fig:Corr} near the point where three string segments are connected on a Y-type junction. This means that instead of focusing on the correlation function on a single string, we deal with correlations between vectors on different strings (which we denoted as $h_i( \Delta l_{c \, g} / \tau)$---recall its definition in Eq. \ref{h}). Its value at zero separation $h_{\Delta l=0}$ is defined by the junction configuration (average angles between strings as in figure \ref{fig:Collision}), which was studied in Sect. \ref{Angles between strings after collision}, and it depends only on the relative velocity of the colliding strings and their tensions.

Combining all results, we can calculate how the junction will grow or shrink on average under the assumption that the network has reached a scaling regime. It is useful to rewrite equation (\ref{LengthChangeEq}) in the following way
\begin{equation}
\begin{gathered}
\label{JunctionDynAv2}
  \Delta \dot{l}_{c \, i}^{(n)} = 1 - \frac{ \mu M_i }{\mu_i \mathcal{M}} h_{0 \, i}\bigg[ \eta_1 \Theta(l_{c \, k} - \Delta l_{c \, k}^{(n)}) \left( \frac{\Delta l_{c \, g}^{(n)}}{\tau} \right) +  \\
  + \eta_2 \Theta(\Delta l_{c \, g}^{(n)}-l_k) \Theta(\xi_c-\Delta l_{c \, g}^{(n)}) \left( \frac{\Delta l_{c \, g}^{(n)}}{\tau} \right)^{2 \chi_i} +\\
  + \Theta(\Delta l_{c \, g}^{(n)}-\xi_c) \bigg]  ,
\end{gathered}
\end{equation}
where $\eta_{1,2}$ are constants that ensure the continuity of the correlation function, while $h_{0} = h_{\Delta l=0}$ is the initial value of the correlation function (see figure \ref{fig:Corr} for a graphical illustration of the meaning of $l_{c \, k}$ and $\xi$). The initial values $h_{0 \, i}$ are calculated from the geometry of strings (see figure \ref{fig:Collision}), taking into account that the initial average angles are given by integration of (\ref{Betas}) -- refer to the examples in figure \ref{fig:Av}.

Let us consider the situation where the tensions of the strings are related by the BPS conditions for cosmic superstrings (\ref{PQtensions}). In the example we will consider below, we will set $\mu_F=1$ and the coupling constant $g_s=0.3$.

\begin{figure}
\centering
\includegraphics[width=0.48\textwidth]{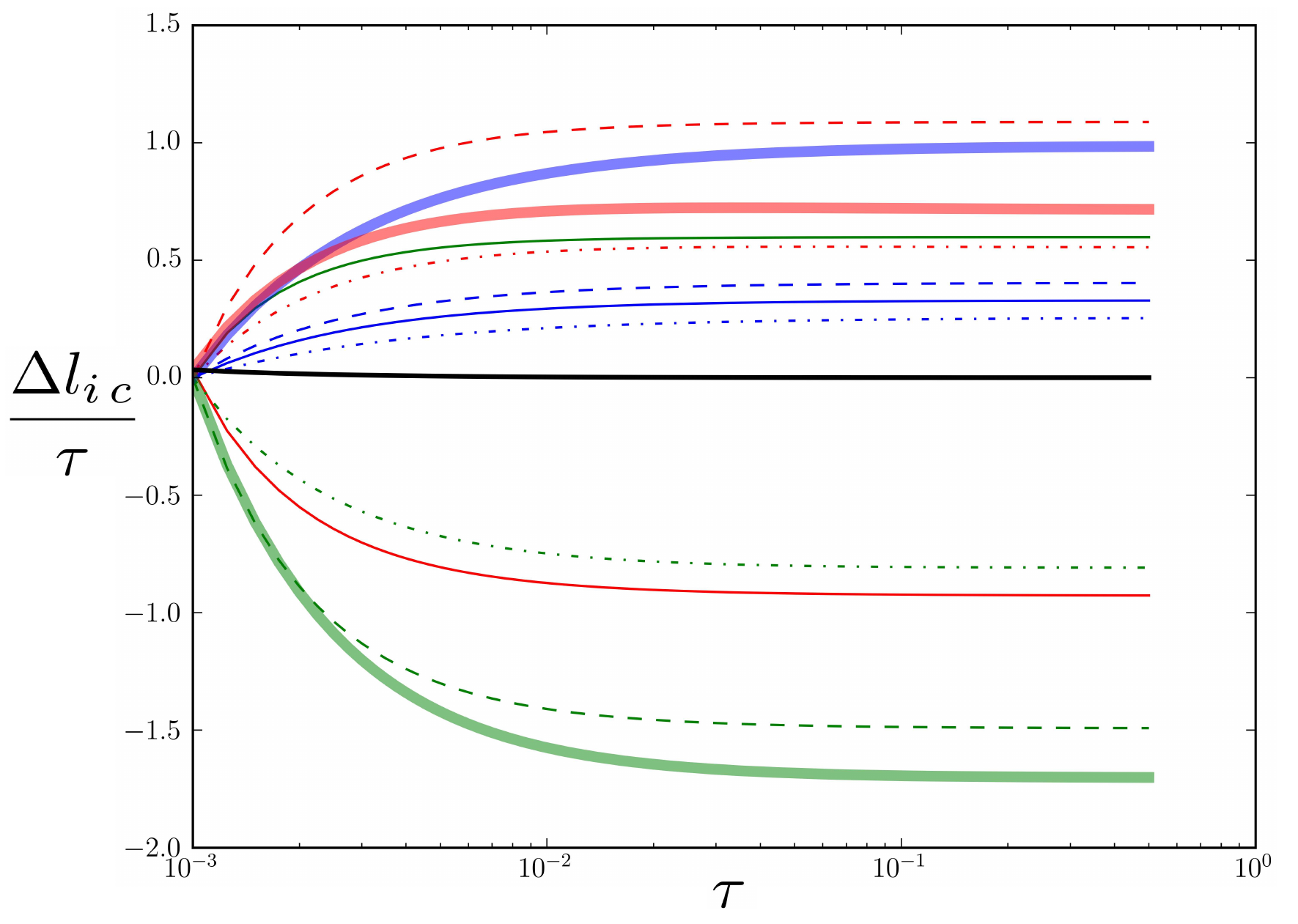}
\caption{\label{fig:SSS} Growth/decrease of length of junction segments $\Delta \ell_i$ for a scaling string network with $\upsilon_i=0.64$, $L_i = 0.3 \tau$, $\xi=\tau$ and tensions defined by (\ref{PQtensions}) with $g_s = 0.3$. The solid thin lines represent the change of the length after collisions of $1$-$2$ strings, the dash-dotted ones to collisions of $2$-$3$ strings and the dashed ones to collisions of $1$-$3$ strings. Thick solid lines show the total length change from the sum of all collisions, while the black line shows the sum of all $\mu_i \Delta \ell_{i \, c}$. Blue lines represent strings of type $1$, red lines strings of type $2$, and green lines strings of type $3$. }
\end{figure}

Each collision between strings of types $1$ and $2$ leads to an increase of the length of string of type $3$. To understand how much this can grow overall in the string network, we need to determine the average initial angle for such collisions. For the string network we fix the macroscopic network parameters according to a scaling regime in the radiation domination period: we have chosen $\upsilon=0.64$, $L=0.3 \tau$, $\xi = \tau$ ($n=1$). As a result, using the average equations (\ref{JunctionDynAv2}) with the ansatz (\ref{PQtensions}) we can obtain the average length growth/reduction for each type of string for all possible collisions. The results are shown in figure \ref{fig:SSS}. The thin solid lines show the resulting length exchange for collisions of $1\equiv (1,0)$ and $2\equiv (0,1)$ strings and the growth of the $3\equiv (1,1)$ string. The blue, red and green colors corresponds to strings of types $1$, $2$ and $3$ respectively. We apply the same treatment for collisions between strings of types $2$ and $3$ (dash-dotted lines in figure \ref{fig:SSS}) and for collisions between strings of types $1$ and $3$ (dashed lines in figure \ref{fig:SSS}). The sum of all length exchanges is shown by the thick lines in figure \ref{fig:SSS}.

\subsection{VOS model for strings with dynamical junctions} \label{VOS model for strings with dynamical junctions}

Having done the relevant preparatory work (sections \ref{Angles between strings after collision} - \ref{Correlation function of incoming components; small-scale structure}), we can return to the VOS model (for a detailed description of it see \cite{MartinsShellard, MartinsShellard2,Book}) and introduce the necessary modifications. In particular, we are aiming to obtain a model where the junction evolution is described by averaged kinematic constraints. Specifically, in this section we are consider a string network with $3$ types of strings, with tensions given by Eq. (\ref{PQtensions}) and with the same value of the string coupling constant $g_s=0.3$ as in the previous section.

To model the network evolution we need to introduce energy exchange terms that describe the averaged dynamics of junctions. For this purpose we use the VOS type of model developed in \cite{AvgoustidisShellard}. The probability of string interactions and their effect were already studied in several previous works \cite{AvgoustidisCopeland, PourtsidouAvgoustidisCopelandPogosianSteer, AvgoustidisShellard3}. Here our main advance is to explicitly introduce the dynamics of junctions, through average growth/decrease lengths.

Our model of multi-tensional string network evolution with dynamical junctions evolving under the above assumptions can be written in the following way, as an extension of \cite{AvgoustidisShellard}:
\begin{equation}
\begin{gathered}
\label{VOSJunctionAllSystem}
  \dot{L}_{c \, i} = \frac{\dot{a}}{a} v_i^2 L_{c \, i} + \frac{1}{2} c_i  v_i - \\
  -\sum_{n,j,k}^3  |\epsilon_{njk}| d_{jk} \frac{  v_{jk}  \Delta l_{c \, i}^{(n)} }{4 L_{c \, j}^2 L_{c \, k}^2} L_{c \, i}^3 , \\
  \dot{v_i} = (1-v_i^2) \left( \frac{k(v_i)}{L_{c \, i}} - 2 v_i  \frac{\dot{a}}{a} \right), \\
   \Delta \dot{l}_{c \, i}^{(n)}  = 1 - \frac{ \mu M_i h_i(\Delta l_{c \, g }^{(n)} / \tau ) }{\mu_i \mathcal{M}},
\end{gathered}
\end{equation}
where
\[
v_{jk}= \sqrt{v_j^2 + v_{k}^2} \,,
\]
\begin{equation}
\begin{gathered}
\label{MomentumK}
  k(\upsilon) = \frac{2 \sqrt{2}}{\pi} (1-\upsilon^2) (1 + 2 \sqrt{2} \upsilon^3 ) \frac{1 - 8 \upsilon^6}{1 + 8 \upsilon^6},
\end{gathered}
\end{equation}
and the parameters $c_i$, $d^{(n)}$ model the probabilities of string interactions and will be defined below.

As mentioned in the introduction, cosmic superstrings can reconnect (interact) with probabilities that can be computed/estimated by the study of fundamental string scattering amplitudes \cite{POLCHINSKI, DaiPolchinski, JacksonJonesPolchinski} and effective field theory on strings \cite{HananyHashimoto}. The probability of a fundamental string $(p'=1,q'=0)$ to interact with a $(p,q)$ bound string state is estimated as \cite{JacksonJonesPolchinski}
\begin{equation}
\begin{gathered}
\label{FProb}
  P_f = \frac{  q^2 v^2 + g_s^2 \left( p - \gamma_v^{-1} \cos \theta \frac{\mu_F}{\mu_{pq}} \right)^2 }{ 8 v \gamma_v^{-1} \sin \theta  \frac{\mu_F}{\mu_{pq}}  },
\end{gathered}
\end{equation}
where $\theta$ and $v$ are the angle and velocity of collision.

The probability of a $(p,q)$ string to interact with another $(p',q')$ string (where $q,q' \geq 1$) is given by expression \cite{JacksonJonesPolchinski, HananyHashimoto}
\begin{equation}
\label{DProb}
  P_d = \text{min} \left[1, \, 1 - \left( 1- P \right)^{q q'}  \right],
\end{equation}
where 
\begin{equation}
\label{Probd}
  P = \frac{\sqrt{g_s}  \text{e}^{2 \sqrt{2/3} \theta / v}}{2 (\pi \theta)^{3/4}} \text{exp} \left[ -\frac{4 \sqrt{\pi} \theta^{3/2}}{g_s} \text{e}^{-4 \sqrt{2/3} \theta / v} \right] \nonumber.
\end{equation}

In order to apply probabilities (\ref{FProb}), (\ref{DProb}) to the VOS model we need to average them over the string network. For this purpose we integrate expressions (\ref{FProb}), (\ref{DProb}) over all possible velocities and angles in the network, taking into account the kinematic constraints arising from the Nambu-Goto action for three strings joining at a junction \cite{AvgoustidisCopeland, PourtsidouAvgoustidisCopelandPogosianSteer}
\begin{equation}
\label{AvProb}
  \mathcal{P} = \frac{1}{N} \int_0^{v_{cr}} \int_0^{\alpha_{cr}} P_{f,d} \;  \text{e}^{ -(v - \overline{\upsilon})^2/\sigma^2_v }   v^2 \sin \theta d \theta d v,
\end{equation}
where $\overline{\upsilon}$ is the rms relative velocity of strings, $\alpha_{cr}$ and $v_{cr}$ are the limits of integration in angles and velocities respectively, defining the region in angle-velocity space for which string collisions lead to junction formation (this region is generally non-trivial so $\alpha_{cr}=\alpha_{cr}(v_{cr}))$.  Note that a large variance is allowed, $\sigma_v=0.5$ \cite{AvgoustidisCopeland}. Finally the normalization factor $N$ is defined as
\begin{equation}
\label{Norm}
  N = \int_0^{1} \int_0^{\pi/2} \text{e}^{ -(v - \overline{\upsilon})^2/\sigma^2_v }   v^2 \sin \theta d \theta d v.
\end{equation}

\begin{figure}
\centering
\includegraphics[width=0.48\textwidth]{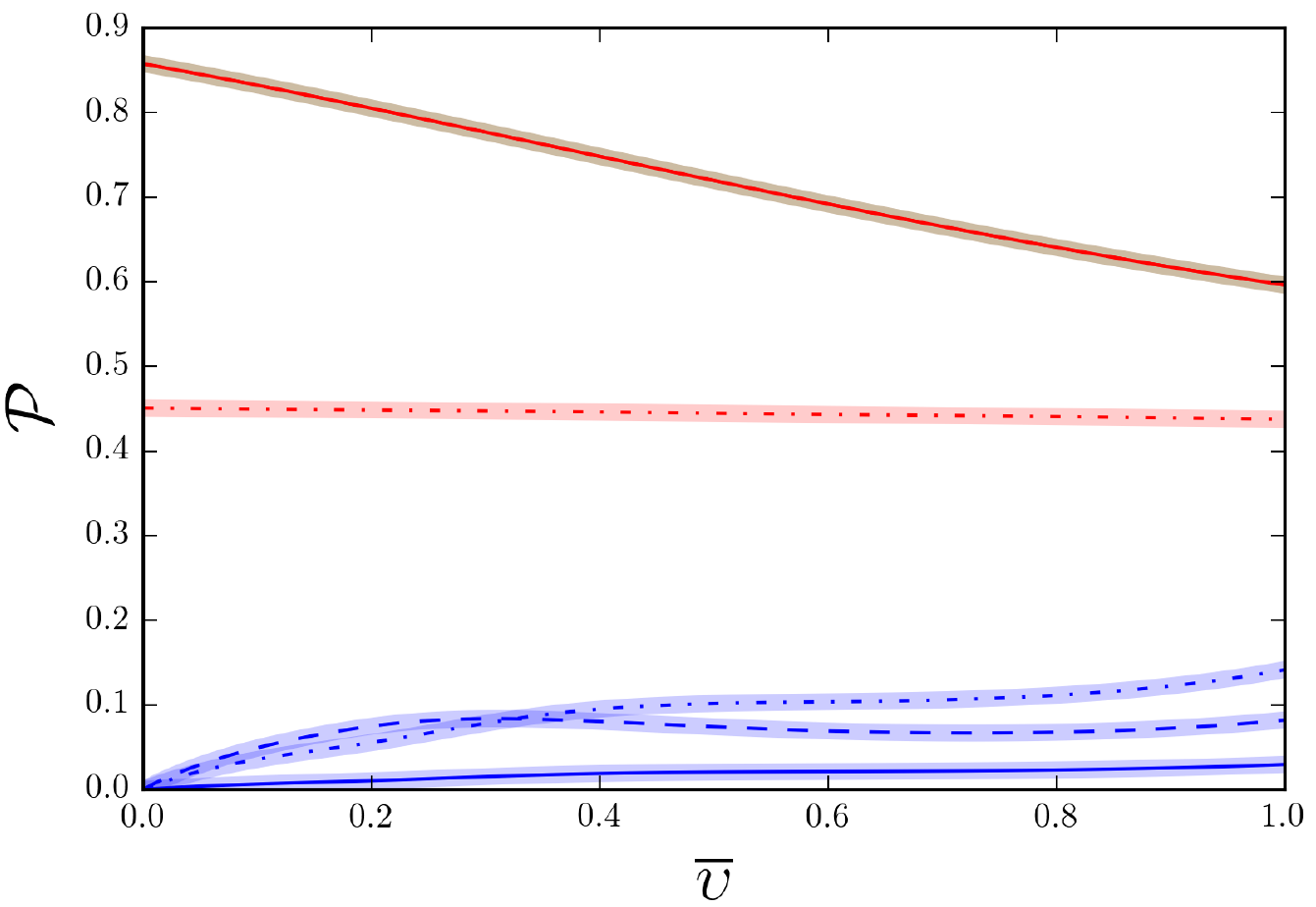}
\caption{\label{fig:ProbabAll} Probabilities of $(p,q)$ string interactions depending on the rms velocity $\overline{\upsilon}$. Blue lines represent the probability of $(1,0)$ string interaction with $(1,0)$ (solid line), $(0,1)$ (dashed line) and $(1,1)$ (dash-dotted line) strings. Red lines show the probability of $(0,1)$ string to interact with $(0,1)$ (solid line) and $(1,1)$ (dash-dotted line) strings. Thick transparent lines present the numerical approximation of these probabilities.}
\end{figure}

The values of the average probabilities as functions of the rms velocity $\overline{\upsilon}$ are shown in figure \ref{fig:ProbabAll}. To speed up the computation of the VOS model scaling solutions, we approximate the integral (\ref{AvProb}) by a table of pre-computed values. These are shown as thick lines in figure \ref{fig:ProbabAll}.

We define the self-interaction coefficients, taking into account probabilities (\ref{AvProb}), in the form
\begin{equation}
\label{Chopping}
  c_i = ( \mathcal{P}_{ii} \, \mathcal{V}_{ii})^{1/3} c,
\end{equation}
where $\mathcal{V}_{ij}$ is a volume factor (the influence of this factor was studied in \cite{PourtsidouAvgoustidisCopelandPogosianSteer}), $\mathcal{P}_{ij}$ is the probability (\ref{AvProb}) for the corresponding string collision, and $c$ is the standard chopping parameter. The power $1/3$ in equation (\ref{Chopping}) comes from simulations of Nambu-Goto networks \cite{AvgoustidisShellard3}. The coefficient of string interactions $d_{jk}$ can be written as 
\begin{equation}
\label{DIJ}
  d_{jk} = ( \mathcal{P}_{jk} \, \mathcal{V}_{jk})^{1/3} d,
\end{equation}
where $d$ is a constant. Here, we will study the case where all volume factors are unity, $\mathcal{V}_{jk}=1$. For a discussion of the effect of these volume factors on network evolution see \cite{PourtsidouAvgoustidisCopelandPogosianSteer}.

\begin{figure}
\centering
\includegraphics[width=0.48\textwidth]{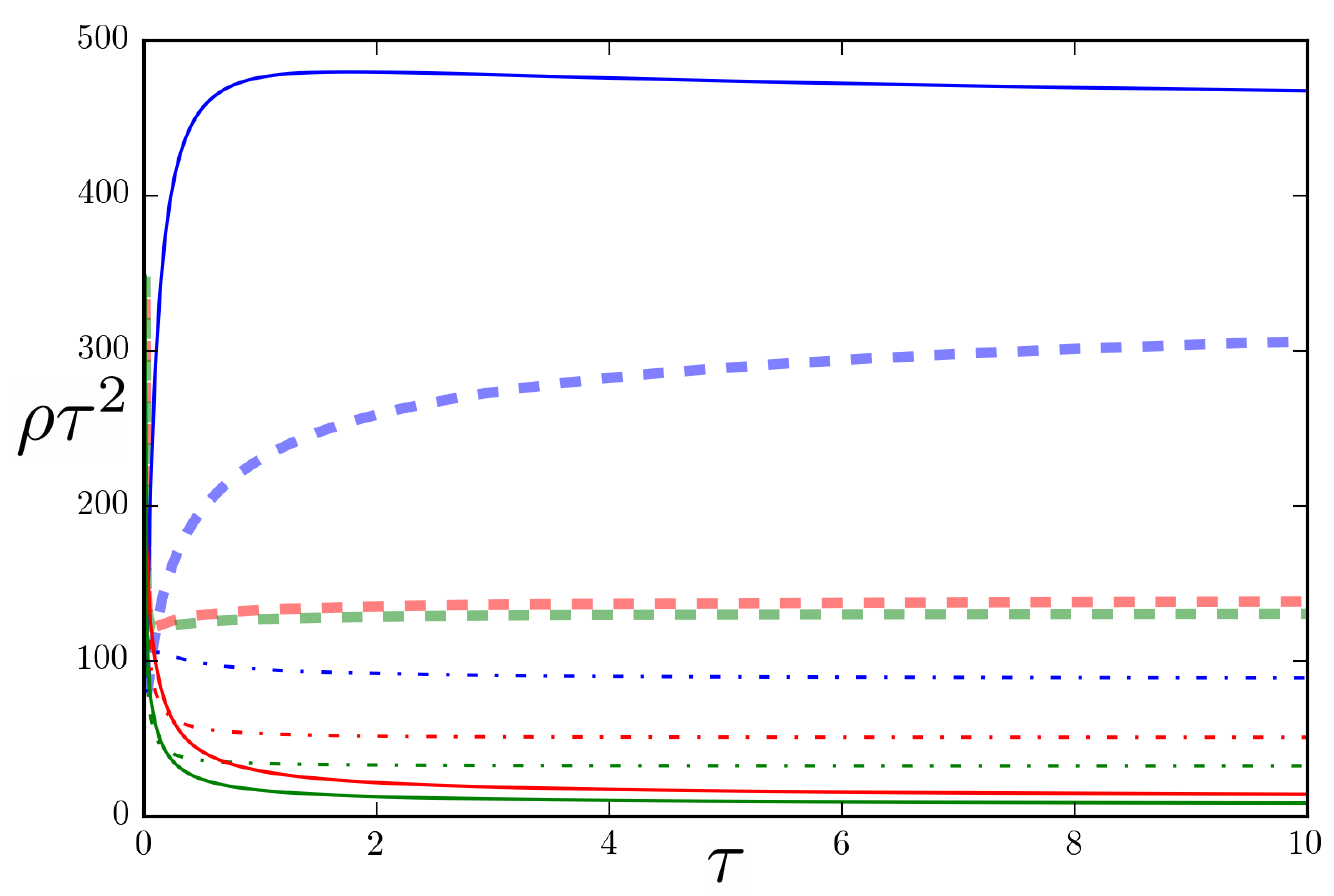}
\includegraphics[width=0.48\textwidth]{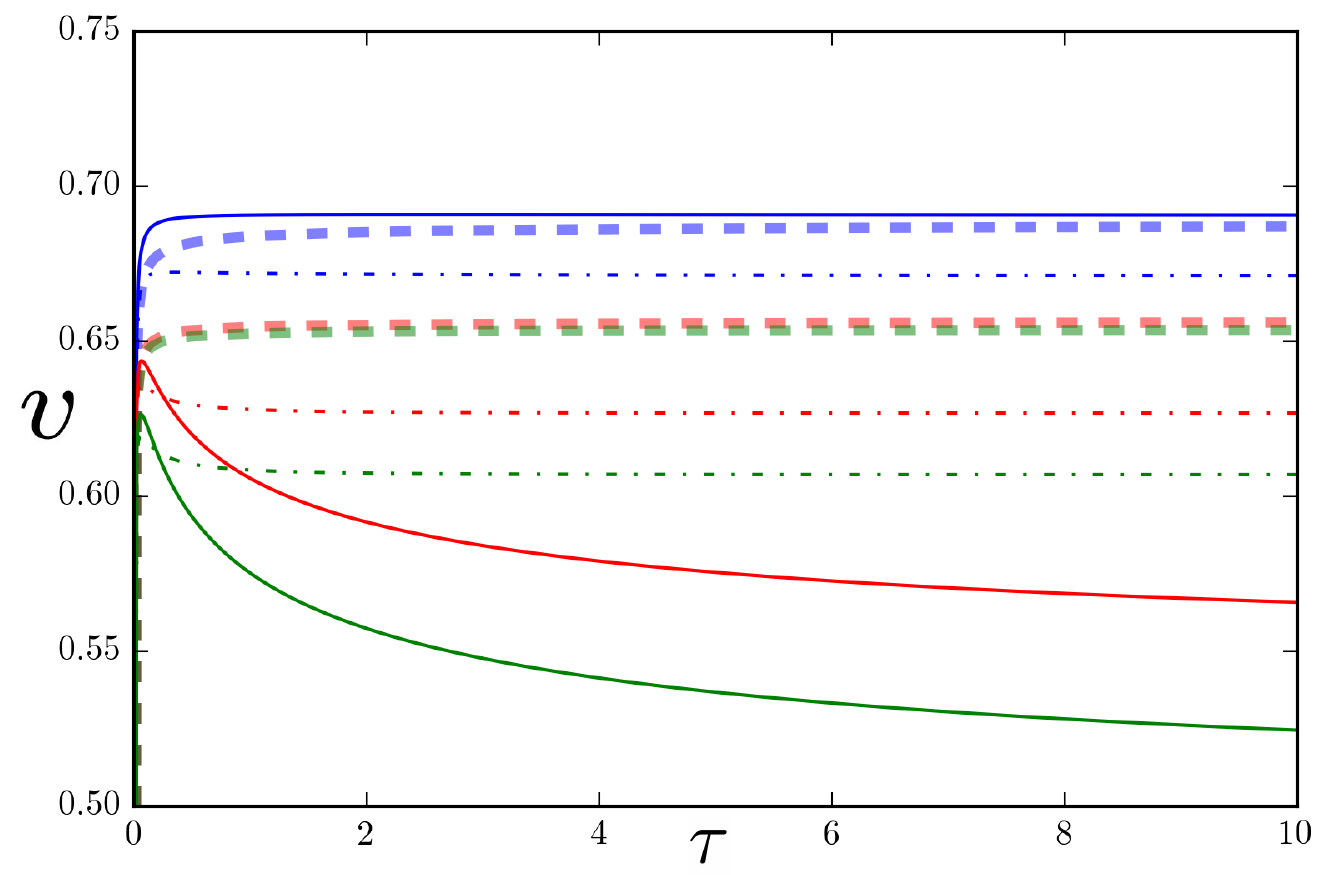}
\caption{\label{fig:Energ} Energies $\rho_i = \frac{\mu_i}{L_i^2}$ and rms velocities of the multi-tensional string network. Green lines correspond to the heaviest strings, red line to the middle ones and blue lines to the lightest strings. Dashed lines correspond to the standard VOS model from \cite{AvgoustidisShellard}, dash-dotted lines to the current model without contribution of probabilities to interact and solid lines represent the full model (\ref{VOSJunctionAllSystem}) studied in this work. Calculations are performed for radiation domination period ($n=1$) with $c=0.27$, $d=0.1$.  }
\end{figure}

Having the form of the parameters $d_{jk}$ and $c_i$ we can solve the VOS model (\ref{VOSJunctionAllSystem}) and look for scaling solutions. Specifically, we will illustrate the model results with the case of the radiation domination epoch ($n=1$) with the following choice of parameters: $g_s=0.3$, $c=0.27$, $d=0.1$, and $k_{k}=0.48$ (see table \ref{TableMeasureChiA} for justification of some values). The result of these calculations is shown in figure \ref{fig:Energ} by the solid lines. It is interesting to compare the result obtained here with the previous approach in \cite{AvgoustidisShellard}, where the energy exchange term was assumed to be $\Delta l_{c}^{(n)} = \frac{1}{2}\sum_{ij} |\epsilon_{nij}| \frac{L_i L_j}{L_i + L_j}$. These calculations are shown in figure \ref{fig:Energ} by the dashed lines. It is worth pointing out that the model developed in \cite{AvgoustidisShellard} predicts an identical evolution for all types of strings if the probabilities of interactions are the same. Meanwhile, the model developed in this work differentiates the evolution of strings with different tensions even if they have identical probabilities to interact. Assuming the probability of interaction for all strings to be unity, we demonstrate the last statement in figure \ref{fig:Energ} by dash-dotted lines.

\section{Conclusions}

In this work we revisited the description of junctions in cosmic string networks, and of their impact in the overall dynamics. We studied kinematic constraints and collision between straight string segments in a generic FLRW background metric. As anticipated, the approach developed in \cite{CopelandKibbleSteer, CopelandKibbleSteer2, CopelandFirouzjahiKibbleSteer} in Minkowski space can readily be generalised to an expanding FLRW metric (as it was done for loops in \cite{FirouzjahiKaroubyKhosraviBrandenberger, FirouzjahiKhoeini-MoghaddamKhosravi}). In particular, for straight string collisions we studied the region in angle-velocity space for which junction production is allowed in an FLRW metric. We demonstrated that the change of the angle-velocity area is caused only by the deceleration of straight strings in an expanding universe.

We thus studied the averaged properties of string collisions within a string network. We computed the string configurations (in particular, the angles between string segments, see figure \ref{fig:Collision}) that should appear on average immediately after string collisions. We argued why junctions should eventually stop growing and also discussed how we can track their dynamics on a macroscopic level. To do so, we connected the equation for junction dynamics with the correlation function along the string, which has been directly measured in Goto-Nambu string simulations. The initial conditions (in the limit of zero separation) for the correlation functions have been obtained from average string configurations just after the collision of strings.

Putting everything together, that is combining the averaged probabilities of string interactions, the correlation functions and the velocity dependent angle configurations, we introduced modifications (which effectively correspond to new energy loss/gain terms) to the VOS model describing the evolution of superstring networks. The resulting VOS model thus includes the dynamics of string junctions. We presented one example where scaling solutions were found for a 3-string toy model in the radiation era; analogous solutions exist for the matter era. Our results on string evolution and the methodology developed here will be useful for further studies of cosmic strings and their observational outcomes.

\begin{acknowledgments}

This work was financed by FEDER---Fundo Europeu de Desenvolvimento Regional funds through the COMPETE 2020---Operational Programme for Competitiveness and Internationalisation (POCI), and by Portuguese funds through FCT---Funda\c c\~ao para a Ci\^encia e a Tecnologia in the framework of the project POCI-01-0145-FEDER-028987. The work of IR is supported by the FCT fellowship SFRH/BD/52699/2014, within the FCT Doctoral Program PhD::SPACE (PD/00040/2012), and by grant CIAAUP-28/2018-BIM, funded by FCT through national funds (UID/FIS/04434/2013) and by FEDER through COMPETE2020 (POCI-01-0145-FEDER-007672). The work of AA was partly supported by an Advanced Nottingham Research Fellowship and an STFC consolidator grant at the University of Nottingham.

\end{acknowledgments}

\section*{\label{APDX}Appendix: On the various definitions of velocity}

In this Appendix we clarify the different velocities used in this work. There are subtle differences among the velocities appearing in our equations, owing to the use of different reference frames in their derivation (both in our own work here and in the already available literature). We have been careful with using the correct velocities in our formulae, but this is not always shown in the notation, which would have otherwise been rather cumbersome.

The computation of the kinematic constraints in \cite{CopelandKibbleSteer, CopelandKibbleSteer2,CopelandFirouzjahiKibbleSteer} and in section \ref{Junction dynamics} of this work is performed in the frame where both of the colliding strings are moving with equal speed towards each other. In particular, the velocity in Section \ref{Junction dynamics} and in Figure \ref{fig:areas} should not be confused with the velocity of collision that can be obtained by the transformation (\ref{VelTransform}).

The results we use for the probabilities of string interactions \cite{JacksonJonesPolchinski, HananyHashimoto} are calculated in the string rest frame in which one of the colliding strings is at rest. Note this is in contrast to \cite{POLCHINSKI, DaiPolchinski}, where the computation is done in the reference frame where the colliding strings have equal speeds. Hence, when we apply the kinematic constraints and intercommuting probabilities for the string network, integrating over all possible angles and velocities, c.f. (\ref{AvProb}), we should use the same reference frame.

Special attention should be given to the transition from microscopic velocities of string segments to averaged velocities for the string network. In this work we used the approach of \cite{AvgoustidisCopeland}, where the integration was carried out under the assumption that, in the string rest frame, the velocity has a Gaussian distribution (\ref{AvProb}). Note, however, that this distribution is not mathematically rigorous and further study is required to demonstrate that it captures the main characteristics of network velocities in string simulations. In addition, in order to reach a more accurate description of the VOS model for multi-tensional string networks (\ref{VOSJunctionAllSystem}), it will be necessary to be able to distinguish between rms velocity, mean velocity, and the average velocity of string collisions. 

\bibliography{junctions}
\end{document}